\documentclass[oldversion]{aa}
\usepackage[switch]{lineno} 
\usepackage{times} 
\usepackage{graphicx}
\usepackage[
singlelinecheck=false 
]{caption}
\usepackage[switch]{lineno} 
\usepackage{xspace} 
\usepackage{epsfig}
\usepackage{amssymb} 
\usepackage{textcomp} 
\usepackage{longtable}
\usepackage{rotating} 
\usepackage{booktabs} 
\usepackage{pdflscape}
\usepackage{amsmath}
\usepackage{natbib} 
\usepackage{tabularx} 
\usepackage{rotating} 
\usepackage{dcolumn}
\usepackage{epstopdf}

 \bibpunct{(}{)}{;}{a}{}{,}

 \newcommand{\kep}{{\em Kepler }}
\newcommand{\xmm}{{\em XMM-Newton }}
\def\gsim{\;\lower4pt\hbox{${\buildrel\displaystyle >\over\sim}$}\,}
\def\lsim{\;\lower4pt\hbox{${\buildrel\displaystyle <\over\sim}$}\,}
\begin{document}
\title{Activity and rotation of the X-ray emitting \kep stars}

\subtitle{}

\author{D. Pizzocaro\inst{\ref{inst1}, \ref{inst2}}, B. Stelzer\inst{\ref{inst3},
\ref{inst4}}, E. Poretti\inst{\ref{inst5}}, S. Raetz\inst{\ref{inst3}}, G. Micela \inst{\ref{inst4}}, A. Belfiore
\inst{\ref{inst1}}, M. Marelli \inst{\ref{inst1}}, D. Salvetti \inst{\ref{inst1}}, A.
De Luca \inst{\ref{inst1}, \ref{inst6}}}

\offprints{}

\institute{ INAF-Istituto di Astrofisica Spaziale e Fisica Cosmica Milano, via E.
Bassini 15, 20133 Milano, Italy\label{inst1}  \\ \email{D. Pizzocaro,
daniele.pizzocaro@gmail.com} \and
 Università degli Studi dell\textquoteright Insubria, Via Ravasi 2, 21100 Varese, Italy\label{inst2}
 \and
 Institut f\"ur Astronomie \& Astrophysik, Eberhard-Karls-Universit\"at T\"ubingen,
 Sand 1, 72076 T\"ubingen, Germany\label{inst3} \and
 INAF - Osservatorio Astronomico di Palermo, Piazza del Parlamento 1, 90134 Palermo,
 Italy\label{inst4} \and
 INAF - Osservatorio Astronomico di Brera, via E. Bianchi 46, 23807 Merate (LC),
 Italy\label{inst5} \and
 INFN - Istituto Nazionale di Fisica Nucleare, Sezione di 
  Pavia, via A. Bassi 6, 27100 Pavia, Italy\label{inst6}\\
} 
\titlerunning{} \authorrunning{D. Pizzocaro et al.}

\date{Received $<$XX-XX-2015$>$ / Accepted $<$XX-XX-2015$>$}

\abstract{ The relation between magnetic activity and rotation in late-type stars
provides fundamental information on stellar dynamos and {\rm angular momentum evolution.} 
Rotation/activity studies found in the literature suffer from inhomogeneity in the
measure of activity indexes and rotation periods. {\rm We overcome this limitation} with 
a study of the X-ray emitting {late-type main-sequence} stars observed by \xmm and \kep. 
We measure rotation periods from photometric variability in
\kep light curves. As activity indicators, we adopt the X-ray luminosity,  
the number frequency of white-light flares, the amplitude of the rotational
photometric modulation, and the standard deviation in the \kep light curves. 
{\rm The} search for X-ray flares in the light curves provided by the EXTraS ({\em Exploring
the X-ray Transient and variable Sky}) FP-7 project {\rm allows us to identify} simultaneous
X-ray and white-light flares. 
{\rm A careful selection of the X-ray sources in the \kep field yields $102$ main-sequence
stars with spectral types from A to M.}  
We find rotation periods for {$74$ X-ray emitting
main-sequence} stars, $22$ of which {without period reported in the previous} literature. 
In the X-ray activity/rotation relation, we 
{\rm see evidence for the traditional distinction of a saturated and a correlated 
part, the latter presenting a continuous decrease in activity towards slower rotators.
For the optical activity indicators the transition is abrupt and located at a period of 
$\sim 10$\,d but it can be probed only marginally with this sample which is 
biased towards fast rotators due to the X-ray selection.} 
We observe $7$ {\em bona-fide} X-ray flares {\rm with evidence for a white-light
counterpart in simultaneous \kep data. We derive an X-ray flare frequency of $\sim 0.15\,{\rm d}^{-1}$, 
consistent with the optical flare frequency obtained from the much longer \kep time-series.}  
}
\keywords{stars: main sequence, rotation, activity, coronae, flares; X-rays}

\maketitle

\section{Introduction}\label{sect:intro} 
Main sequence stars are characterized by {\rm
radiative} emission processes, such as high-energy (UV and X-ray) emission, flares
{and} enhanced {optical} line emission (\ion{Ca}{\sc ii}, H$\alpha$), 
{collectively} referred to as `magnetic
activity', as they are ascribed to processes involving magnetic fields in the stellar
corona, chromosphere and photosphere. Magnetic activity is allegedly the result of
internal magnetic dynamos, arising from the combination of stellar differential
rotation and convection in the sub-photospheric layers.

The {understanding} of
stellar magnetic activity is of capital importance, since it is a fundamental
diagnostics for the structure and dynamics of stellar magnetic fields, and gives
crucial information on 
the dynamo mechanism 
responsible for their {\rm existence}. Beside this, the high-energy
emission associated with magnetic activity  has a fundamental role in the evolution
of the circumstellar environment and on the composition and habitability of planets.

As stated above, rotation is one of the key ingredients
of stellar dynamos. In a feedback mechanism, the coupling of the rotation itself with
the magnetic field 
determines the spin evolution of stars. This is true both in the pre-main sequence
phase, in which the angular momentum of the accretion disk is transferred to the
central forming star through the magnetic field, and in the main-sequence phase,
because of the momentum loss due to magnetized winds ejected by the star. Exploring
the relation between the stellar magnetic activity and the {star's rotation rate} is 
thus an {efficient} way to gather information on stellar dynamos.\\
\indent The connection
between stellar rotation and magnetic activity has been studied in many works, since
the seminal papers by \cite{1966ApJ...144..695W} and \cite{1967ApJ...150..551K}. A
fundamental contribution was given in the work by \cite{1972ApJ...171..565S}, the
first to interpret the activity-rotation relation as a consequence of a dynamo
mechanism. Since then, many authors (e.g., \citealt{1981ApJ...245..671W},
\citealt{1989ApJ...344..907D}, {\citealt{1985ApJ...292..172M}, \citealt{2003A&A...397..147P}}) 
have focused on the relation between
the stellar rotation and specific chromospheric and coronal activity indicators. From
these works, a bimodal distribution emerged for the rotation-activity relation: for
rotation periods longer than a few days (depending on the spectral type of the star,
or the equivalent parameters of stellar mass or colour), the activity decreases with
the rotation period; for shorter periods, a saturation regime is reached, in which
the activity level is independent {\rm of} the rotation period. 
Since rotation is an ingredient of the dynamo
mechanism, a correlation between rotation and activity is intuitively reasonable. The
origin of the saturation, instead, has not been ascertained. 
{\rm It may be} due to a change in the behaviour of the dynamo, or {due to  
limits to the coronal emission} because the stars run out {of} the available 
surface to accomodate more active regions \citep{1995A&A...294..715O} or because the high
rotation rate causes centrifugal stripping of the stellar corona
(\citealt{1999A&A...346..883J}).

X-ray emission is a very good proxy
for stellar activity: The X-ray activity of a star is 
{\rm the result of} 
magnetic reconnection in the stellar corona, 
{\rm an abrupt change in the configuration of the magnetic field} 
determining the release of non-potential energy
stored in the magnetic field lines. Rotation-activity studies found in the literature
{\rm typically} refer to collections of X-ray data obtained from various instruments,
and rotation periods measured with different techniques, combining ground-based
photometric measurments with $v\,\sin i$ spectroscopic techniques. The limitations
due to the use of inhomogeneous data sets can now be overcome by combining 
{rotation periods from the {\em Kepler} optical light curves to 
X-ray data obtained with {\em XMM-Newton} 
which has observed $\sim 1500$ objects in the \kep field of view.}

In the present work we study the relation between the rotational properties
and the magnetic activity of the X-ray emitting main-sequence field stars observed {with} the
{\em Kepler} mission \cite{} comparing various indicators for activity (X-ray and UV
activity,
white-light flaring rate, flare amplitude, {\em Kepler} light curve amplitude and
standard deviation) and rotation (rotation period and Rossby number;
\citealt{1984ApJ...279..763N}).

The sample selection is described in Sect.
\ref{sect:sample}. The procedure used to evaluate the physical parameters of all
stars in the sample is described in Section \ref{sect:params}. In Section
\ref{sect:kepler} we describe the techniques used to determine the rotation period
{and photometric activity diagnostics.} 
In Section
\ref{sect:act}, several indicators of stellar activity are analysed: the X-ray
luminosity, the ultraviolet excess in the Spectral Energy Distribution (SED), the
white-light and X-ray flaring activity. The results are {discussed in 
Sect.~\ref{sect:results} and conclusions are presented in Sect.~\ref{sect:conclusions}.}

\section{Sample selection} \label{sect:sample} 

A proper sample selection is crucial
in order to obtain a reliable picture of the activity-rotation relation. Many
rotation-activity studies have been performed on inhomogeneous samples of stars for
which X-ray data had been collected from a set of various databases and using a
combination of spectroscopic and photometric techniques for rotation measurements.

We aim to the highest homogeneity in the determination of the rotation period
and in the characterization of the activity indicators, first of all the X-ray
activity. To this end, we {focus in this work on} X-ray emitting stars detected
by \xmm with light curves from {\em Kepler}. We perform a positional match between
the 3XMM-DR5 catalogue \citep{2015arXiv150407051R} 
and the \kep Input Catalogue (KIC, \citealt{2011AJ....142..112B}),  
and then remove non-stellar objects and objects with uncertain photometry.

\begin{figure}[!htb] \begin{center}
\includegraphics[width=9.2cm]{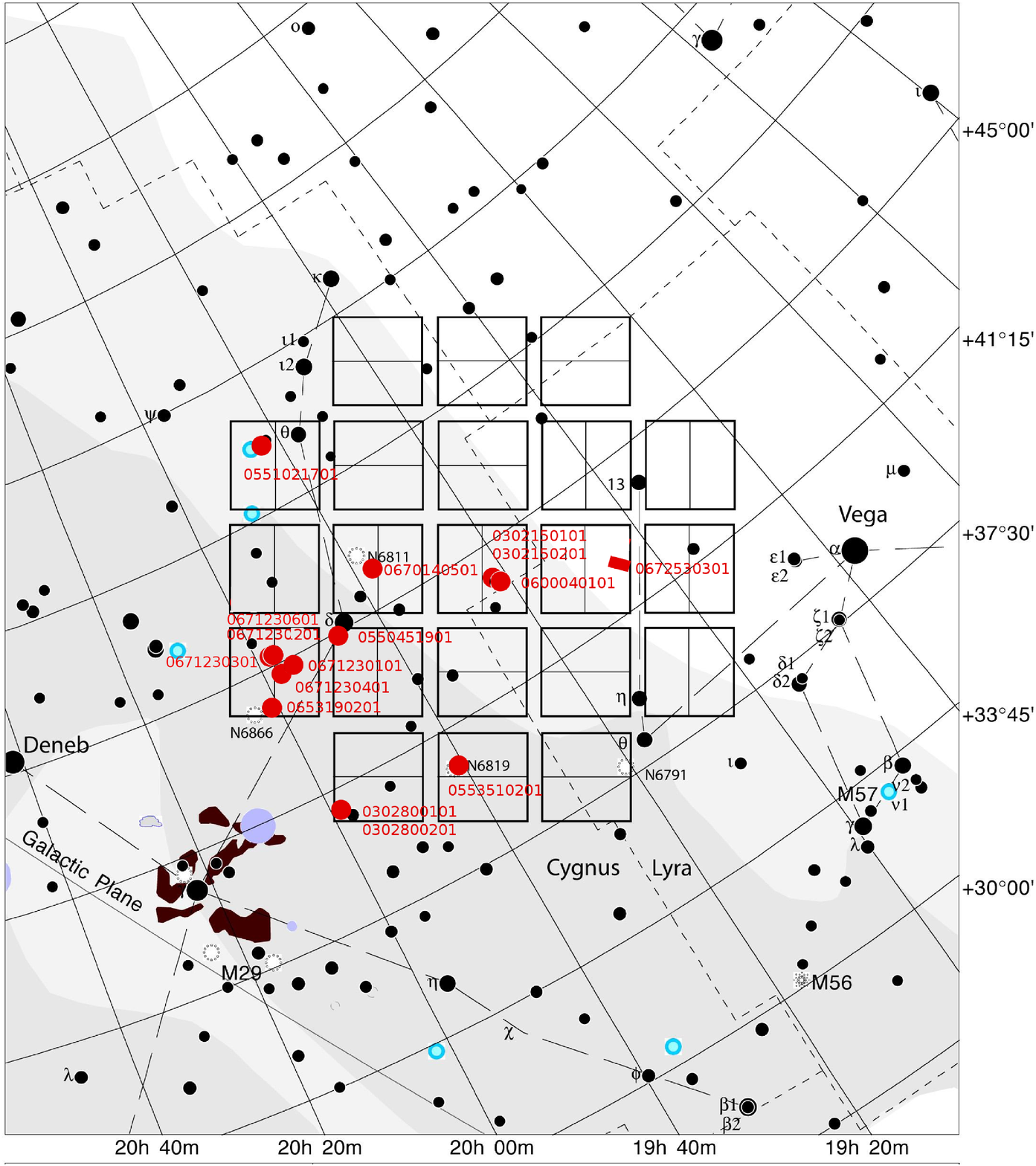}
\caption{The field of
view (FoV) of the {\em Kepler} mission is respresented by the solid black squares. It
is centered at RA=$19°\,22\,40.0\,$, DEC=$+44°\,30\,00.0\,$, between the Cygnus and
the Lyra constellation. The FoVs of the sixteen 3XMM-DR5 observations we analysed
(Table \ref{tab:obs}) are reported as red circles; the red numbers represent the
3XMM-DR5 unique observation ID (OBS\_ID). The FoVs of some {\em XMM-Newton}
observations are totally or partially overlapping.} 
\label{fig:obs} 
\end{center}
\end{figure}

{The 3XMM-DR5 catalogue includes data from} $16$ {\em XMM-Newton} 
pointings within the field of view (FoV) of the {\it Kepler} 
mission (see Table 1). Their sky position is shown in  Fig. \ref{fig:obs}. The {\em
Kepler} FoV covers $\sim 105$ square degrees; the $16$ {\em XMM-Newton} observations
which fall inside this FoV {\rm cover only} $\sim 2\%$ of {\rm that} area.
\begin{table*} 
\begin{center} 
\caption{\xmm observations {from the 3XMM-DR5 Catalogue in
the \kep FoV}. Next to observation ID
(col.\,1), pointing center (cols.\,2 and\,3), observing date (cols.\,4) and exposure
time (col.\,5), the flux sensitivity limit is given (col.\,6) 
calculated as described in Sect.~\ref{subsect:sample_completeness}.}  
\label{tab:obs} 
\begin{tabularx}{\textwidth}{lcccrr} 
\hline
OBS\_ID&RA(J2000)&DEC(J2000)&Obs. date&Exp. time & $f_{\rm X,lim}$ \\
 & & & & [ks] & [${\rm erg / cm^2 / s}$] \\ \hline 
0302150101&$19\,21\,05.99$&$+43\,58\,24.0$&2005-10-10&16.9&$1.46\cdot10^{-14}$\\
0302800101&$19\,59\,28.28$&$+40\,44\,02.0$&2005-10-14&22.5&$3.35\cdot10^{-14}$\\
0302800201&$19\,59\,28.28$&$+40\,44\,02.0$&2005-10-16&18.8&$1.43\cdot10^{-14}$\\
0302150201&$19\,21\,05.99$&$+43\,58\,24.0$&2005-11-14&16.9&$6.27\cdot10^{-15}$\\
0553510201&$19\,41\,17.98$&$+40\,11\,12.0$&2008-05-18&26.9&$2.8\cdot10^{-15}$\\
0551021701&$19\,41\,51.86$&$+50\,31\,01.8$&2008-11-08&11.7&$2.27\cdot10^{-15}$\\
0550451901&$19\,47\,19.42$&$+44\,49\,40.8$&2008-11-20&17.9&$1.16\cdot10^{-14}$\\
0600040101&$19\,21\,11.41$&$+43\,56\,56.2$&2009-11-29&58.3&$8.35\cdot10^{-15}$\\
0653190201&$20\,01\,40.00$&$+43\,52\,20.0$&2010-06-23&15.9&$4.02\cdot10^{-14}$\\
0670140501&$19\,36\,59.90$&$+46\,06\,28.0$&2011-05-10&31.9&$2.44\cdot10^{-15}$\\
0672530301&$19\,05\,25.90$&$+42\,27\,40.0$&2011-06-05&29.1&$1.93\cdot10^{-15}$\\
0671230401&$19\,58\,00.01$&$+44\,40\,00.0$&2011-06-09&85.9&$2.53\cdot10^{-15}$\\
0671230301&$19\,58\,00.01$&$+45\,10\,00.0$&2011-06-23&83.9&$3.37\cdot10^{-15}$\\
0671230101&$19\,55\,59.98$&$+44\,40\,00.0$&2011-09-25&81.9&$2.53\cdot10^{-15}$\\
0671230201&$19\,55\,59.98$&$+45\,10\,00.0$&2012-03-26&104.5&$1.84\cdot10^{-15}$\\
0671230601&$19\,55\,59.98$&$+45\,10\,00.0$&2012-03-26&2.7&$1.10\cdot10^{-13}$\\ 
\hline
\end{tabularx} 
\end{center} 
\end{table*} 

The KIC positional error ($\sim 0.1^{\prime\prime}$) is negligible with respect to the 
error in the 3XMM-DR5 position ($\lesssim
4\,^{\prime\prime}$ at $3\,\sigma$). We select all the 3XMM-DR5 unique\footnote{The 3XMM-DR5
Catalogue contains a row for each X-ray detection. For a certain ``unique'' source
(the astrophysical object), many detections (referred to different observations) of
that object may be available.} sources that have a positional match (columns SC\_RA,
SC\_DEC in 3XMM-DR5) with one or more KIC objects within a radius given by three
times their 3XMM-DR5 individual ($1\,\sigma$) positional error (column SC\_POSERR). We calculate
the average probability of a chance association between a 3XMM-DR5 source and a KIC
source in each {\em XMM-Newton} field as $P=1-e^{\pi \mu r^2}$, where $\mu$ is the
numerical density of KIC sources in the field and has a typical value of
$10^{-4}\,{\rm sources/arcsec^2}$, and $r$ is three times the average position error
of the 3XMM-DR5 source. We obtain an average probability of chance association of
$\sim0.8\,\%$. This {translates to} $\sim 0.14$ chance associations per {\em
XMM-Newton} field, i.e. $\sim 2$ spurious matches in the whole sample. 

We select only
the objects with a detection significance in 3XMM-DR5 (column DET\_ML) greater than $6$
(probability of a spurious detection $<0.025$) in at least one EPIC instrument in the
energy band $8$ ($0.2-12.0\,${\rm keV}), removing the others from the sample ($4$
objects removed). We reject the $21$ sources {that} are classified as ``extended"
in 3XMM-DR5 (flag EP\_EXTENT$>6$). The resulting sample consists of $145$ matches.
There are no multiple associations, that is associations between a KIC object and two
or more 3XMM-DR5 objects, or viceversa.

Within this sample, we aim to
identify genuine stars, removing {(1) galaxies and (2) stars that} possibly suffer 
confusion
(stars which are not resolved by {\em Kepler}) and contamination from nearby bright
sources. {\rm To this end,} for each KIC object with a 3XMM-DR5 match, we search for
a classification in the SIMBAD database\footnote{http://simbad.u-strasbg.fr/simbad/} \citep{2000A&AS..143....9W} and we inspect the optical and infrared
images, when available, using the Aladin service \citep{2000A&AS..143...33B,
2014ASPC..485..277B}. 
The fit of the Spectral Energy Distribution (SED) with a model for the stellar
photospheric emission is also useful to identify galaxies in the sample (objects
whose SED cannot be fit by a stellar model). {\rm Moreover, it allows us} to
determine the value of the stellar fundamental parameters. The methods to remove
non-stellar counterparts to the X-ray sources are explained in more detail in the
{following}.

\subsection{Spectral energy distribution (SED)} \label{subsect:sample_sed}

\subsubsection{Multi-wavelength photometry} \label{subsubsect:sample_sed_photo} 

We extract the IR
(2MASS $J$, $H$, $K$), optical (SLOAN $g$, $r$, $i$, $z$, Johnson $U$, $B$, $V$, 
{\em Kepler} $K_{\rm p}$) and  UV (GALEX $FUV$, $NUV$) photometry 
for all stars of {our} sample from the KIC
\citep{2011AJ....142..112B}. Many sources lack photometry in one or more of these
bands. We exclude from the sample the KIC objects for which no 2MASS IR photometry is
available ($2$ objects), since the wavelength range covered by 2MASS is crucial in
order to perform a reliable spectral {classification}.
 
The visual inspection
of the optical images via the Aladin server {indicates} that some stars suffer
contamination from {brighter objects located nearby. In such cases, even if they 
are recognised as distinct objects in the KIC,ß the photometry of the 
fainter one is expected to be significantly contaminated by the brighter one.} 
Other stars suffer confusion between unresolved objects, 
i.e. they are not resolved in the KIC survey, 
but can be recognised as two objects {through} visual inspection of the optical
image. When the object is resolved in the UCAC-4 Catalogue
\citep{2012yCat.1322....0Z}, we replace the KIC photometry for these confused objects
with the photometry provided by UCAC-4 for the brightest one. If no UCAC-4 photometry is available, we
remove the object from our sample ($2$ objects).

According to
\cite{2012ApJS..199...30P}, the SLOAN bands photometry reported in the KIC presents a
significant systematic error, as observed comparing the KIC photometry in the bands
$g$, $r$, $i$, $z$ with the photometry reported in the {\em Sloan Digital Sky Survey}
(SDSS-DR8) in the same bands, available for $\sim 10\%$ of the {\rm full} KIC.
\cite{2012ApJS..199...30P} give semi-empirical formulas to correct these systematics
(their Eqs. 1, 2, 3, 4). We apply these corrections to the KIC SLOAN bands photometry
of all the stars of our sample. The photometry for all the stars in our sample is
reported in Table \ref{tab:photometry}.

\subsubsection{SED fitting}\label{subsubsect:sample_sed_fitting}

After the SEDs have been compiled, we {fit them with photospheric 
BT-Settl models \citep{2012RSPTA.370.2765A}
within} the {\em Virtual Observatory SED
Analyser} (VOSA, \citealt{2008A&A...492..277B}). 
The parameters of {the BT-Settl} models are:
$A_{\rm V}$, $\log{g}$, $[Fe/H]$ and $T_{\rm eff}$. Due to parameter degeneracy it is not
possible to determine all of them from the SED fit. Therefore we adopt for each star
for $A_{\rm V}$, $\log{g}$ and $[Fe/H]$ values taken from the literature 
(see Sect.~\ref{sect:params}), leaving $T_{\rm eff}$ as the only free parameter.
We thus obtain, for each object, a best fit $T_{\rm eff}$ under the hypothesis of a
stellar model. As the uncertainty on the effective temperature we assume the typical
dispersion of the five best fit values obtained in VOSA for each object ($\pm
200\,${\rm K}).

From the shape of their SED,  combined with the visual inspection of the Aladin
images, sixteen objects are recognised as galaxies, and we remove them from our sample.

The whole sample selection procedure eventually results in a sample of $125$ 3XMM-DR5
unique sources with a reliable  stellar counterpart in {\em Kepler}, with a SED that
is well-fitted by a stellar photosphere model.
%
%
%

\subsection{X-ray flux limit} \label{subsect:sample_completeness} 

The sample of stars used
in the present work has been selected based on the X-ray emission observed by 
{\em XMM-Newton} {and as such is biased towards X-ray active stars.} A
fundamental step in order to understand the results presented {in this work}  
is, {therefore}, to provide an evaluation of the 
{X-ray flux limit} 
of the sample. To this
end, we first produce the EPIC PN\footnote{The EPIC ({\em European Photon Imaging
Camera}) consists of three CCD detectors, two MOS and one PN, located at the focus of
the three grazing incidence multi-mirror X-ray telescopes which constitute the main
instrument onboard \xmm.} sensitivity map for each \xmm observation of
Table~\ref{tab:obs}, for the energy range $0.2-2.0\,{\rm keV}$ 
using the task \texttt{esensmap} of the \xmm Science Analysis Software (SAS).
The sensitivity map provides in each point of the FoV the limiting count rate needed to
have a $3\,\sigma$ detection of a point source. We measure the limiting count rate in
the centre of the FoV. 
{We convert these numbers into flux sensitivity limits ($f_{\rm X,lim}$) using 
WebPIMMS\footnote{http://heasarc.gsfc.nasa.gov/cgi-bin/Tools/w3pimms/w3pimms.pl} 
assuming an APEC model with plasma temperature $0.86\,{\rm keV}$ 
(see Sect.~\ref{subsubsect:act_xray_spec}), abundance $0.2$ in solar units and the galactic 
hydrogen column density provided by \cite{2005A&A...440..775K}. 
The resulting numbers for $f_{\rm X,lim}$ are reported in the last column of Table~\ref{tab:obs}.} 
{These values give an idea of the lower limit for the X-ray flux in these observations 
(median is $\sim 6\cdot10^{-15}\,{\rm erg/cm^2/s}$), although we caution that 
the sensitivity varies strongly (up to about one order of magnitude)
from the centre to the outer region of the FoV.}

\section{Fundamental stellar parameters} \label{sect:params} 

For $119$ stars out of
$125$ we found literature values for effective temperature ($T_{\rm eff}$), visual
absorption ($A_{\rm V}$) surface gravity ($\log g$), and metallicity (${\rm [Fe/H]}$) (in
\citealt{2014ApJS..211....2H} and \,\,\citealt{2016arXiv160609149F}).
\cite{2014ApJS..211....2H} {\rm present} a compilation of literature values for
atmospheric properties ($T_{\rm eff}$, $\log g$ and ${\rm [Fe/H]}$) derived from different
observational techniques (photometry, spectroscopy, asteroseismology, and exoplanet
transits), which were then homogeneously fitted to a grid of {\rm isochrones from the
{\em Dartmouth Stellar Evolution Program} \citep[DSEP;][]{2008ApJS..178...89D}}, for a set
of $\sim200,000$ stars observed by the {\em Kepler} mission in Quarters $1-16$.
\cite{2016arXiv160609149F} present a systematic spectroscopic study of $\sim50,000$
{\em Kepler} stars performed using the LAMOST telescope \citep{2012RAA....12..723Z}.
We compared the parameters obtained by
\cite{2014ApJS..211....2H} and \cite{2016arXiv160609149F} {\rm for our sample}, in particular the
effective temperatures, {and we found that they} 
agree very well within the error bars. When available,
however, we prefer the values obtained from \cite{2016arXiv160609149F} through the
spectroscopic analysis, {since in that work} 
the stellar parameters
{have been evaluated}
using the same method for all the stars in their sample, while
\cite{2014ApJS..211....2H} present a collection of values obtained in several works
using different techniques.

Six stars in the sample 
do not have $T_{\rm eff}$ reported in \cite{2014ApJS..211....2H} or
\cite{2016arXiv160609149F}. 
For these $6$ stars we adopt as $T_{\rm eff}$  the effective temperature of the best
fit of the individual SED with a model of the stellar photosphere, as described in
Sect. \ref{subsect:sample_sed}. These stars are flagged with the `VOSA' flag in column $12$ of Table $7$. 
Two out of these $6$ stars have values for $A_{\rm V}$, ${\rm [Fe/H]}$ and
$\log g$ in \cite{2014ApJS..211....2H} or \cite{2016arXiv160609149F}, and we use
them in the SED fit. For the others, we adopt the median values of the distribution
of $A_{\rm V}$ in our sample ($0.38$), ${\rm [Fe/H]}_{\odot}$ ($0.0$) and $\log{g}$ ($4.0$). 
{Similarly, for the additional $15$ stars that have no literature value for $A_{\rm V}$
we adopt the median of $0.38$\,mag.}
{The adopted values for the spectral parameters of all $125$ stars
are listed in Table~\ref{table:params}.}

\begin{figure}
\begin{center}
\includegraphics[width=10.0cm]{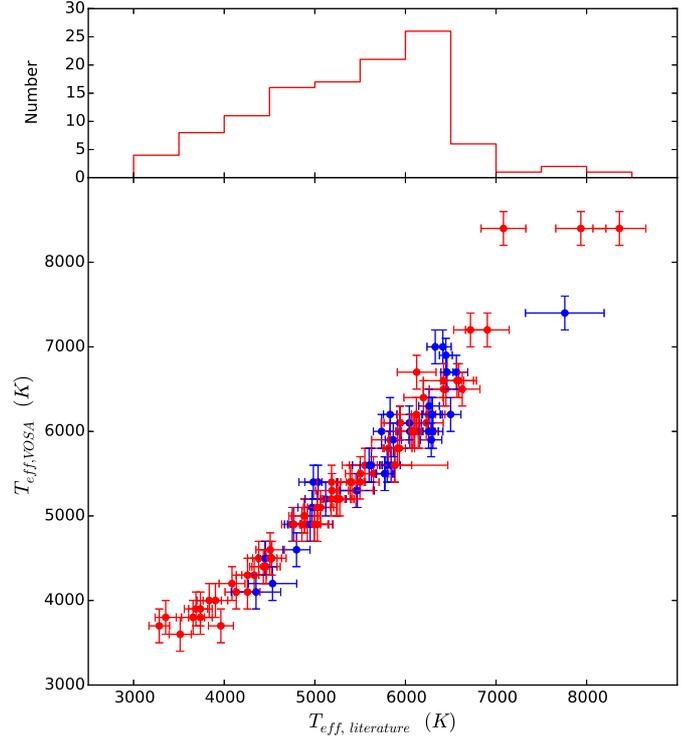} 
\caption{In the upper panel, the distribution of the $T_{\rm eff}$ of the $119$ stars in our
sample with $T_{\rm eff}$ taken from the literature \citep{2014ApJS..211....2H,
2016arXiv160609149F}. In the lower panel, the comparison between the $T_{\rm eff}$
drawn from the literature and the ones obtained with the SED fitting is reported for
the same sample of stars. The colours represent the original work in which the
$T_{\rm eff}$ was derived.  Blue: \cite{2016arXiv160609149F}; red:
\cite{2014ApJS..211....2H}. 
}
\label{fig:teff}
\end{center}
\end{figure}

In order to validate our SED fitting procedure 
we compare the obtained $T_{\rm eff}$ with the ones from the above-mentioned
literature sources. The distribution of $T_{\rm eff}$ and the comparison between the
$T_{\rm eff}$ obtained from the SED fitting and the values in the literature are
reported in Fig.~\ref{fig:teff}. Within the error bars, the agreement is {for most
stars} very good, but the error bars of the values obtained by SED fitting are
typically larger than the error bars in \cite{2014ApJS..211....2H} and especially in
\cite{2016arXiv160609149F}. This {justifies} our decision to adopt, when available, the
literature values for $T_{\rm eff}$. Given the effective temperature, 
we assign a spectral type {to each star} according to Table 5 in
\cite{2013ApJS..208....9P}. {These values are given in Table~\ref{table:params}.}

\begin{figure}[!htb] 
\begin{center}
\includegraphics[width=9.3cm]{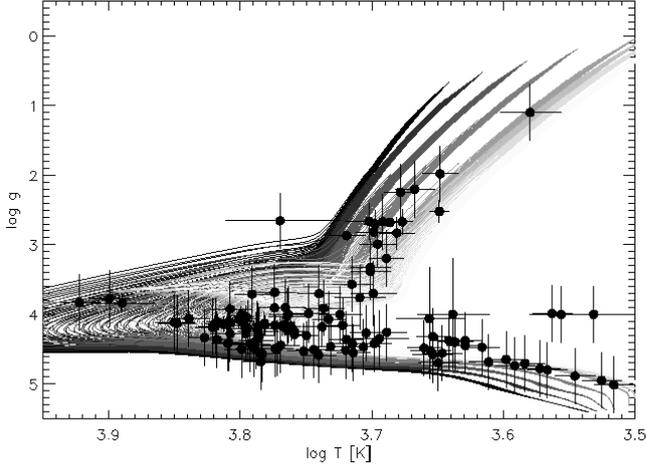} \caption{$\log{g} - \log{T_{\rm eff}}$
space with the set of isochrones from DSEP and the stars of our sample (black
circles) overplotted. Each stack of isochrones (lines in different grey-shades)
corresponds to a given value of $[Fe/H]$ ($-2.0$,
$-1.5$, $-1.0$, $-0.5$, $0.0$, $0.2$, $0.3$, $0.5$ from dark to pale).} 
\label{fig:dsep}
\end{center}
\end{figure}

\subsection{Distance, mass and bolometric luminosity} \label{subsect:params_dsep} 

{Fig.~\ref{fig:dsep} shows our stellar sample together with the DSEP isochrones on 
the $\log g$ vs $\log{T_{\rm eff}}$ plane.}
For each
star, we estimate 
{absolute $J$ band magnitude}, mass ($M$) and bolometric luminosity ($L_{\rm bol}$) by
{projecting its position and
the related uncertainties onto the isochrones (along the vertical axis) in Fig.~\ref{fig:dsep}.} 
{For this task we make use of DSEP isochrones in the range of $1.0$ to $9.5\,${\rm Gyr}  
and for each star, we select the set of isochrones 
 corresponding to the ${\rm [Fe/H]}$ value closest to {\rm the observed value}.} 
There are a few stars the position of which on the $\log g$ vs $\log{T_{\rm eff}}$ 
plane is not consistent, 
within the error bars, with any of the DSEP isochrones. We mark them with a flag
(`DSEP outlier') in Table \ref{table:params}.

{Inverting the parallaxes provided in the Gaia-DR2 \citep{2018A&A...616A...1G}
we obtain the distances to our targets. While the overall quality of the Gaia-DR2 data is 
excellent, the mission is too complex to achieve optimal calibrations with less than two years 
of observations. As a result the Gaia-DR2 still contains many spurious astrometric solutions 
\citep{2018A&A...616A..17A}. To remove putative problematic solutions we cleaned our dataset 
using the filters defined by \citet[][appendix C, equations C-1 and C-2]{2018A&A...616A...2L}. 
Furthermore we used additional quality indicators of the solutions 
(astrometric\_excess\_noise\,$>$\,0, astrometric\_gof\_al\,$>$\,5) to discard other potential 
outliers. Two stars of our sample have no Gaia parallax while the solutions for further 
$60$ stars might not be reliable and were filtered out. For these $62$ targets we derive 
the photometric distance from the comparison between the absolute magnitude and the 
observed $J$ band magnitude. Comparison of the photometric and astrometric distances for 
the whole sample shows overall good agreement. For $5$ targets with reliable Gaia distances 
we found the $J$ band distances to have lower uncertainties. In these cases we decided 
to adopt the $J$ band distances. To summarize, throughout this work we use the Gaia distances 
for $58$ stars and the photometric $J$ band distance for the remaining stars.}  
Distance, mass and bolometric luminosity are reported for all {$125$ stars} in 
Table~\ref{table:params}.

\begin{figure}[t] 
\begin{center}
\includegraphics[width=13.0cm,angle=270]{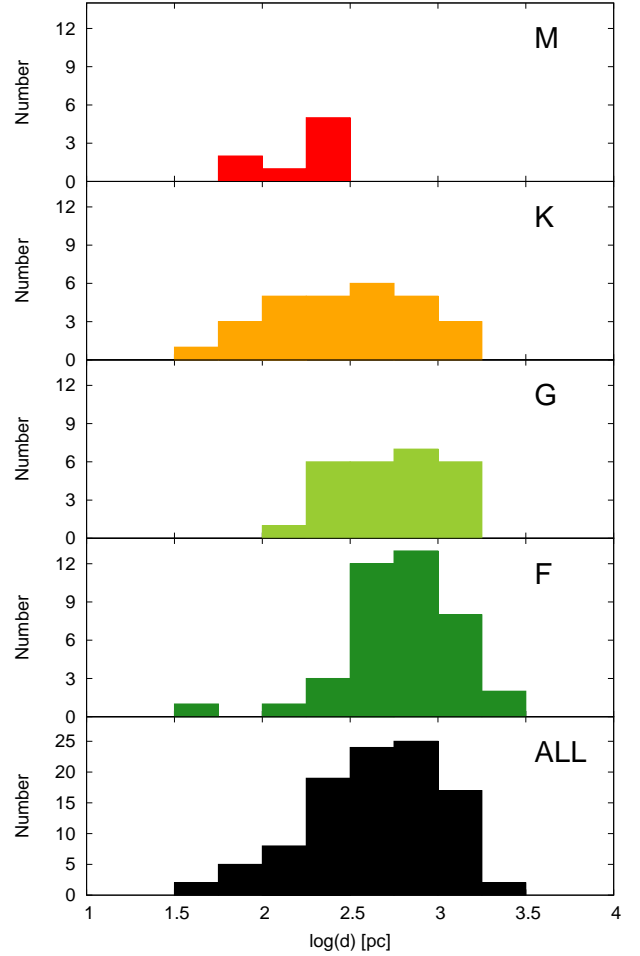}
\caption{Distribution of the {adopted} distances 
for the individual spectral types (from top to bottom: M, K, G, F) and for the whole sample {of the main-sequence stars}
({bottom} panel).} 
\label{fig:dist_d} 
\end{center} 
\end{figure}

We classify the stars in our sample as main-sequence
stars or giant stars according to their $\log g$. In the $\log g$ vs $\log{T_{\rm eff}}$
space, the stars in our sample {can be separated} into two groups, corresponding to the dwarf
branch and to the giant branch, respectively (see Fig.~\ref{fig:dsep}). On this
basis, we consider as main-sequence stars all the stars with $\log g\geq3.5$, and as
giants all the stars with $\log g<3.5$. With this criterion we find $19$
giant stars. 
{The four stars without $\log g$ value in the literature can not be assigned to 
these groups, and we eliminate them from the sample. These stars have a flag (*) in column $12$ of Table $7$.}
This work is focused on main-sequence stars. {Therefore,} in the following, if not
differently declared, we consider only the sample of the $102$ main-sequence stars.
The distribution of {their adopted distances} is plotted in Fig. \ref{fig:dist_d}. 

\section{Analysis of {\em Kepler} light curves} \label{sect:kepler} 

The brightness
modulation in the optical light curve due to the presence of inequally distributed
spots on the rotating stellar surface enables the measurement of the stellar rotation
period. Rotation periods, together with {\rm several} photometric activity
diagnostics, are extracted from the {\em Kepler} light curves following the procedure
described by \cite{2016MNRAS.tmp.1060S}. The methods developed therein for the
analysis of M dwarf K2 mission lightcurves can readily be applied to the data from
the main \kep mission, and we briefly resume the steps in Sect.~\ref{subsect:kepler_spot}. 

To obtain a clean sample of stars for the rotation-activity study, periods
originating from mechanisms different from the rotational brightness modulation due
to starspots must be identified and removed. In particular, our sample covers a broad
range of $T_{\rm eff}$ in the Hertzsprung-Russell diagram, including the intersection
of the main-sequence with the classical instability strip, where stellar pulsations
are expected. Therefore, the light curves of some stars deserved a more detailed
study in order to ascertain if pulsation (but also binarity) could be the real cause
of the observed variability. We perform this study by means of  the Fourier
decomposition \citep{1994A&A...285..524P} and the iterative analysis of the whole
frequency spectrum \citep{1971Ap&SS..12...10V}, 
{described in Sect.~\ref{subsect:kepler_variability}.}

\subsection{Search for rotational periodicity} \label{subsect:kepler_spot}

Briefly, the analysis consists of the following steps, implemented in an iterative
procedure, which is described in detail by \cite{2016MNRAS.tmp.1060S}: 
(1) period search with standard time-series analysis techniques (autocorrelation function
[ACF] and Lomb-Scargle 
[LS] periodogram), {(2) boxcar smoothing of the lightcurve and subsequent subtraction of
the smoothed from the observed lightcurve, effectively removing the 
rotational modulation,} 
(3) identification of `outlier' data points in the `flattened' lightcurve obtained from
step (2) through $\sigma$-clipping. The `outliers' comprise both instrumental
artefacts and flares. The latter ones are identified by imposing three criteria: (i)
a threshold ($3\sigma$) for the significance of the flare data points above the
`flattened' curve, (ii) criterion (i) applies to at least two consecutive such data
points, and (iii) the maximum bin of the flare has at least a factor two higher flux
than the last bin defining the flare. For a detailed description of the procedure used 
in each step of the analysis chain see \cite{2016MNRAS.tmp.1060S}.\\ 
\indent After the removal of the `outliers'
we repeat the period search, but in practice the applied methods are so robust that
the cleaning of the lightcurves does not alter the result. What does change after the
cleaning is the amplitude of the rotation cycle, when measured between maximum and
minimum flux bin. However, as we describe below in
Sect.~\ref{subsect:kepler_activity}, the preferred characterization of the spot cycle
amplitude involves diagnostics that are little affected by the small fraction of
`outlier' data points. We perform this analysis for each {\em Kepler} observing
Quarter independently. This allows us to cross-check the results by comparing the
periods obtained from different Quarters, as explained in 
Sect.~\ref{subsect:kepler_rot}. 

The determination of the rotation period for a given star 
proceeds in the following way. We use the routines
\textsc{A\_CORRELATE} and \textsc{SCARGLE} in the IDL environment\footnote{IDL is a
product of the Exelis Visual Information Solutions, Inc} to {generate} the ACF and LS
periodogram. We work independently on the ACF and LS periodogram series. For each
{\em Kepler} observing Quarter, we inspect visually the ACF and LS periodogram
generated by our procedure, and search for a signal of periodic variability (a sharp
peak at a certain frequency, possibly followed by the related harmonics). If absent,
we reject the Quarter from the analysis.
In the ACF periodogram, we visually choose the highest peak of the series, which
corresponds to the period of the modulation; this is generally the first one, or the
second one if the light curve shows a double-humped pattern (as described below). In
the LS periodogram, we likewise choose the highest peak.

For the majority of stars, the dominating periods derived with both techniques (ACF
and LS) are consistent with each other. Deviations regard lightcurves with a
double-humped shape. In this case, the light curve shows a double peak in at least
some of the \kep observing Quarters, while in others a single peak pattern may be
observed. We interpret these features as the signature of two groups of spots at a
roughly antipodal position on the photosphere of the star, one of which may
occasionally disappear. In this case, the ACF periodogram shows generally a first
peak (corresponding to {\rm half the period}) that has a lower amplitude with respect
to the second one, and this pattern repeats for the peaks corresponding to integer
multiples of the first-peak period and of the second-peak period, respectively. In
the Lomb-Scargle periodogram, the peak corresponding to the shorter period often has
a larger amplitude than the longer-period peak. The interpretation of these patterns
as the effect of two groups of spots on the photosphere allows us to interpret the
longer period of the two, corresponding to the double of the other, as the true
rotation period of the star.

We obtain for each star $N_Q$ periods, where
$N_Q$ is the number of Quarters {in which the star shows} a periodic signal. 
This number varies
individually for the stars of our sample, independently for the ACF and the LS
method. We adopt as the star's period the median value of the periods obtained
with the ACF method from the individual quarters.

\subsection{Non-rotational periodicity} \label{subsect:kepler_variability}

For some of the stars in our sample
periodic variability is identified 
{that} 
can not be explained
with a simple spot pattern. In such cases the Fourier decomposition and the
evaluation of the whole frequency spectra were very helpful. The frequency spectra of
the genuine spotted stars are characterized by the harmonics of the rotational period
and by low-frequency peaks due to the activity and rotational cycle-to-cycle period
variations due to the shift in the spots' latitude and to stellar differential
rotation, and by some rotational cycle-to-cycle variation in the shape and amplitude
of the lightcurve due to the variation of the area and shape of starspots. Other
kinds of periodicity present different patterns.\\
\indent We perform a dedicated
anlysis of the light curves of the stars which do not show a clear
variability pattern in order to classify their non-rotational variability. We find
one multiperiodic star showing the high-frequency pulsational regime of $\delta$ Sct
stars and three multiperiodic stars showing the low-frequency regime of $\gamma$ Dor
stars.
In the light curves of $2$ stars 
the amplitudes of the even Fourier harmonics are much larger than those of the odd
ones, as necessary to  fit both the sharp minima and the large maxima shown by
contact binaries. For $3$ stars it remains ambiguous if the variability is due to
star spots or due to orbital motion in a contact binary; 
{\rm another} $2$ stars display non-periodic variability that we could not classify. 
 Finally, $5$ stars are likely rotational variables but displaying more than one
 period possibly indicating a binary composed of two spotted stars 
or a complex pattern with uncertain period. 
These $16$ stars are removed from the sample considered for the rotation-activity
relation. A summary of the number of main-sequence stars in  each
variability class for spectral type is given in Table \ref{table:var}.
\begin{table*} 
\begin{center} \tiny{\caption{Classification of the
main-sequence stars in our sample according to the type of variability observed in
the {\em Kepler} light curves.} \label{table:var}
\begin{tabularx}{\textwidth}{lccccccr} \hline 
SpT&Main Sequence (MS) Stars&With
rotation period&No period&Ecl. binaries&Other binaries&Pulsators&Unclear\\ \hline \\ 
 All & 102 & 74 & 2 & 10 & 5 & 3 & 3 \\
 A  &   3  &  2 & 0 &  0 & 0 & 1 & 0 \\ 
 F  &  37  & 24 & 2 &  5 & 2 & 1 & 1 \\
 G  &  26  & 19 & 0 &  1 & 0 & 1 & 1 \\ 
 K  &  28  & 21 & 0 &  4 & 3 & 0 & 0 \\ 
 M  &   8  &  8 & 0 &  0 & 0 & 0 & 1 \\ 
\hline 
\end{tabularx}
}
\end{center} 
\end{table*}

\subsection{Final sample of spotted stars} \label{subsect:kepler_rot} 

From the analysis
described above, we {can derive a rotation period for $74$ stars in our sample, i.e. 
these stars are inhomogeneously spotted.}  
For the {subsequent} analysis, we also 
compute the Rossby number, defined as $R_0=P_{rot}/\tau_{conv}$, i.e. {the} 
ratio between the rotation period {and the} convective turnover time. 
The convective turnover time is the
characteristic time of circulation within a convective cell in the stellar
sub-photosphere, and not directly observable. We calculate it from $T_{\rm eff}$
using Eq. $36$ in \cite{2011ApJ...741...54C}, which is valid in the range
$3300\,${\rm K}$\,\lesssim\,T_{\rm eff}\,\lesssim\,7000\,${\rm K}. All but two of the
stars in our sample are within this range of $T_{\rm eff}$. The rotation periods and
Rossby numbers {for all ``spotted" stars} are listed in Table \ref{tab:rot}.
{Our subsequent photometric variability analysis is limited to this sample.}

\subsection{Photometric activity diagnostics} \label{subsect:kepler_activity}

From the analysis of the {\em Kepler} light curves we also obtain various diagnostics
for the stellar photospheric activity: the light curve amplitude and the standard
deviation of the light curve.

The light curve amplitude is the photometric
difference between the maximum and the minimum of the rotationally-modulated light
curve, determined  (in a light curve cleaned from flares) by the contrast between
spotted and unspotted photosphere, {combined with the inhomogeneous distribution of spots 
which causes the spot coverage of the observed stellar hemisphere to vary during 
the rotation.} Analogous to \cite{2016MNRAS.tmp.1060S} and for
consistency with the previous literature we decided to define the spot cycle
amplitude as the range between the $5th$ and $95th$ percentile of the observed flux
values in a single {rotation} cycle ($R_{\rm var}$, see \citealt{2013ApJ...769...37B}), 
and we adopt the modified definition of $R_{\rm var}$ introduced by \cite{2013ApJ...775L..11M},
$R_{\rm per}$, which is the mean of the $R_{\rm var}$ values measured individually on
all observed rotation cycles, expressed in percent. 
Cutting the upper- and lower-most $5$\,\% of the data points is another way of
removing the `outliers', such that no difference between the $R_{\rm var}$ values
obtained from the original light curve with the $5th$ and $95th$ percentile and from
the cleaned lightcurve is expected. We have verified this by comparing the $R_{\rm
var}$ values extracted from the light curves `cleaned' from flares and outliers and
the $R_{\rm var}$ values obtained from the original light curves. 

The second activity diagnostic extracted from \kep light curves {that} we
use to characterize the variability during the spot cycle is the standard deviation
of the whole light curve ($S_{\rm ph}$), and the average of the standard deviations
computed for time intervals $k\cdot P_{\rm rot}$, with $k$ integer, first defined by
\cite{2014JSWSC...4A..15M}. \cite{2014JSWSC...4A..15M} 
have shown for a sample of $22$ stars that roughly after five rotation cycles the
full range of flux variation is reached. Therefore, we compute $S_{\rm ph}$ and $\langle
S_{\rm ph,k=5}\rangle$ for the stars in our sample. The standard deviation of the light
curve was used as a proxy of magnetic activity also by \cite{2015ApJS..221...18H}, in
a study on the activity of two solar-like \kep stars.

Finally, we have computed the standard deviation of the `flattened' light
curves ($S_{\rm flat}$), measured on the light curves cleaned from flares and outliers,
and from which the rotational cycle has been removed (see description at the
beginning of this Section). \cite{2016MNRAS.tmp.1060S} have established this
parameter as an indicator for low-level unresolved astrophysical variability such as
{small unresolved flares} and/or small and fast-changing spots for their {sample of
nearby M dwarfs observed in the K2 mission.} 

\begin{figure}
\begin{center}
\includegraphics[width=7.0cm, angle=270]{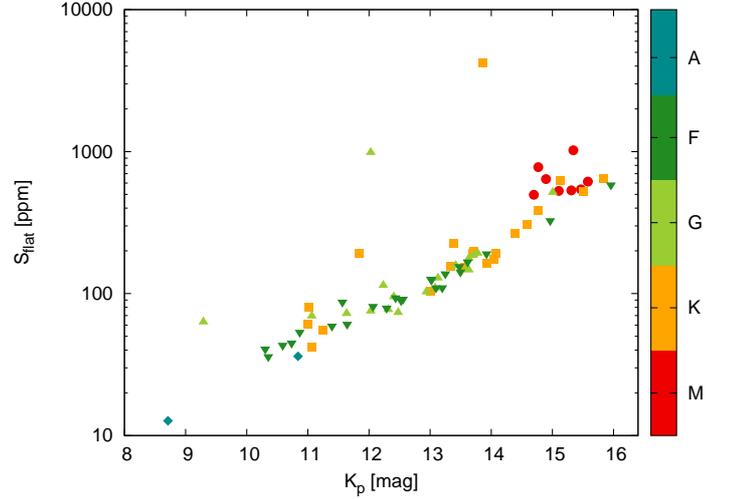}
\caption{{Standard deviation of the `flattened' lightcurve ($S_{\rm flat}$) vs {\em Kepler}
magnitude. Shown is for each star the minimum of $S_{\rm flat}$ from all quarters
of observation.} 
}
\label{fig:sflat_kp} 
\end{center} 
\end{figure}

All these photometric activity diagnostics are listed in Table~\ref{tab:rot}. 
{It is worth noting here that the noise level in the `flattened' lightcurve is 
expected to depend on the brightness of the star. In fact, Fig.~\ref{fig:sflat_kp}
shows a correlation between the {\em Kepler} magnitude and $S_{\rm flat}$. As can be seen
in Fig.~\ref{fig:sflat_kp}, 
the apparent brightness of our sample stars is on average larger for earlier spectral types. 
As a consequence, the noise tends to be smaller in those stars. This must be
taken into account in the analysis of other kinds of variability; see e.g. 
Sect.~\ref{subsect:results_kepler}.} 

\section{X-ray and UV activity} \label{sect:act} 

As our sample {\rm is X-ray selected, every star in the sample has an associated X-ray detection. 
%
{In the main-sequence sample, $36$ stars}
have also been detected in one or both of the UV energy bands (Near
Ultra-Violet, NUV, and Far Ultra-Violet, FUV) of GALEX. 
{In this section we} analyse the X-ray and UV
luminosity, and the corresponding `activity indexes', defined as the ratio between
the luminosity in the {\rm high-energy band} and the bolometric luminosity of the
star, as indicators of the stellar magnetic activity.

\subsection{X-ray data analysis} \label{subsect:act_xray} 

\subsubsection{Source X-ray luminosity}\label{subsubsect:act_xray_cr} 

We calculate the X-ray luminosity of each star in the sample in the
{energy band $0.2-2.0\,${\rm keV}} (`soft' energy band, 3XMM-DR5 catalogue energy band
$6$). 
{This range of energies corresponds approximately to the energy
bands used in previous studies of the rotation-activity connection 
that were mainly based upon {\em ROSAT} data (e.g.
\citealt{2011ApJ...743...48W}).} 

In the 3XMM-DR5 catalogue, there are many objects (`sources') that
have been observed more than once: so, for a certain `source', there can be many
`detections'. Each {row} of the catalogue -- which represents an individual
detection of a certain source -- contains a set of parameters referred to the detection
itself, and a set of parameters averaged on all the detections available for that
source. The catalogue provides the fluxes expected for a power law emission model for
both the `detections' and the `source'. However, the power law model, to which the
fluxes given in 3XMM-DR5 refer, is not appropriate {for describing} the X-ray emission
{of} late-type stars. Therefore, we re-calculate the X-ray flux in the bands
{$0.2-2.0\,${\rm keV} from} the EPIC PN count rate (or the
MOS, if PN is not available), using the HEASARC online tool WebPIMMS, in which we
assume for the X-ray emission of the stars a thermal APEC model
\citep{2001ApJ...556L..91S}. 3XMM-DR5 gives the `detection' count rate for each of the
three EPIC ({\em European Photon Imaging Camera}) instruments (PN, MOS1 and MOS2) on
board the {\em XMM-Newton} mission, and the associated uncertainty, but it does not
give the `source' count rate. As described above, the `detection' count rate is
different from the `source' count rate when the source has multiple detections in the
catalogue. This happens for $11$ sources in our sample. So, we calculate the X-ray
flux for these objects  rescaling one of the individual `detection' count rate for
the factor 
\begin{equation} 
\mathbf{f=\frac{SC\_FLUX_{0.2-2.0}}{DET\_FLUX_{0.2-2.0}}}
\end{equation}
where $\mathbf{SC\_FLUX_{0.2-2.0}}$ and $\mathbf{DET\_FLUX_{0.2-2.0}}$ are
respectively the `source' and `detection' flux provided by 3XMM-DR5. 

The APEC model
in WebPIMMS requires as input parameters the hydrogen column density ($N_{\rm H}$), the
metallicity and the temperature of the emitting plasma. We {\rm estimate} the
hydrogen column density for each source from the visual absorption $A_{\rm V}$ following
\cite{1989ApJ...345..245C}, as 
\begin{equation} 
N_{\rm H}=A_{\rm V}\,{\rm [mag]}\cdot 1.79\cdot 10^{21}\,{\rm cm^{-2}}.  
\end{equation}
We adopt an average abundance of $0.2$ in
solar units, which is a typical value for the coronae of X-ray emitting late-type
stars (see e.g. \citealt{2012MNRAS.419.1219P}). 
We calculate the count-to-flux conversion factor with WebPIMMS. To
establish an operational $kT$, we fit the spectra of the brightest X-ray sources in
XSPEC, {and assume for all stars the $kT$ value in the WebPIMMS grid ($kT=0.86\,{\rm keV}$) 
which is closest to the average $kT$ obtained from the spectral fits ($kT=0.83\,{\rm keV}$).}
The spectral analysis of the brightest sources is described in 
{the following.} 

\subsubsection{Spectral analysis} \label{subsubsect:act_xray_spec} 

We select the subsample of stars
for which we have at least $200\,$ events in the source extraction region in the
three EPIC instruments together (PN+MOS1+MOS2). Excluding the stars classified as
eclipsing or contact binaries, we have a sample of $19$ stars.
For each one, we extract the source and background spectrum in the energy
band $0.3-10.0\,${\rm keV}, and perform a joint  spectral analysis with XSPEC 12.8.1
\citep{1996ASPC..101...17A} of the spectra of all EPIC instruments available for that
source. We fit the spectra with an absorbed one-temperature APEC model
(\texttt{phabs*apec}) or an absorbed two-temperature APEC model
(\texttt{phabs*(apec+apec)}). We set the coronal abundance to the typical value of
$0.2$ in solar units (see above) to reduce the number of free parameters and avoid
degeneracy. 

Two stars show a significant parameter degeneracy in their spectrum, so we
remove them from the analysis. For the stars for which the model requires two
temperatures, we calculate an average temperature weighted on the flux of the two
APEC components. The results of the best fit for each of the $17$ stars, all
characterized by a null-hypothesis probability  $>0.3\,\%$, are reported in Table
\ref{tab:spectra}. We calculate the average $kT$ over the $17$ stars
($<kT>=0.83\,${\rm keV}).

With the parameters given in Sect.~\ref{subsubsect:act_xray_cr},
we obtain from WebPIMMS the expected flux for each source. We then compare these fluxes
with the fluxes obtained from the spectral analysis, for each of the $17$ sources for
which the spectral fit in XSPEC was performed. For each {of these stars} 
we calculate the ratio
between the XSPEC flux and the flux from WebPIMMS. For PN, MOS1 and MOS2, we found an
average ratio of respectively {$0.91, 0.89, 1.00$ in the soft energy band}
$0.2-2.0\,${\rm keV}. We correct the fluxes obtained with WebPIMMS for the faint ($<200$
counts) sources by multiplying them with this factor. This correction is meant to
obtain from WebPIMMS a flux that is, on average, as close as possible to the actual flux
obtained from a spectral fit. 

{In the sample of $17$ bright stars, $kT$ ranges from 
$0.49\,{\rm keV}$ to $1.3\,{\rm keV}$. 
This range is typical for active stars, and the actual plasma 
temperatures of the bulk of the faint stars is likely in the same range. 
This span of temperatures introduces an error in the flux calculated 
with the average $kT$, allowing fluxes which are up to $\sim2\%$ lower or $\sim15\%$ higher 
than the one calculated from the average temperature of the bright stars. If the 
$17$ X-ray-brightest sources 
are not representative  
of our whole sample, 
these errors may be somewhat larger for the faint stars.} 

From the corrected fluxes we calculate the X-ray luminosity using the
distances derived in Sect.~\ref{subsect:params_dsep}. For each source we calculate the X-ray
activity index as 
\begin{equation} 
AI_X=\frac{L_{\rm X}}{L_{\rm bol}} .
\end{equation} 
The {distributions} of the X-ray luminosity and of the
{corresponding} activity index in the {soft energy band ($0.2-2.0\,${\rm keV}) are} 
reported in Fig.~\ref{fig:lx_ai}. The individual 
{X-ray luminosities} are listed in Table~\ref{tab:xray}.

\begin{figure}
\begin{center}
\includegraphics[width=10.0cm,angle=270]{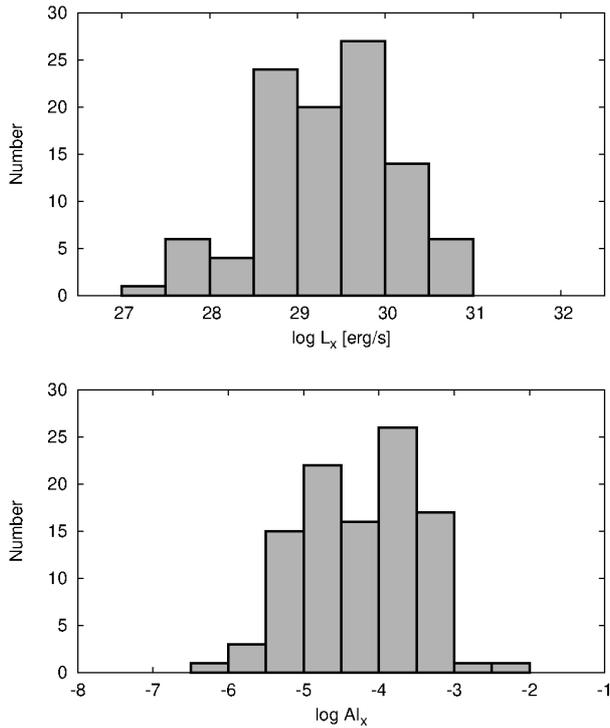}
\caption{Distribution of the X-ray luminosity {and of the X-ray} activity index in the 
energy band $0.2-2.0\,${\rm keV} for {the $102$ main-sequence stars.}}
\label{fig:lx_ai}
\end{center}
\end{figure}

\begin{table*}
\begin{center} 
\caption{Best fit {\rm spectral} parameters for the $17$ brightest
X-ray sources (more than $200$ counts) {\rm identified as rotational variables; parameters
are for} a \texttt{phabs*apec} or \texttt{phabs*(apec+apec)} model.} 
\label{tab:spectra}
\begin{tabularx}{\textwidth}{lcccccc} 
\hline 
KIC\_ID & SpT$^{(a)}$ & Counts$^{(b)}$ & Model & ${\rm N_H}$ & kT & EM  \\ 
        &             &                &       & [${\rm cm^{-2}}$] &  [{\rm keV}] & [${\rm cm^{-3}}$] \\ 
\hline 
5112508&M&520&phabs*apec&$\sim 0$&$0.83$&$1.31\cdot10^{52}$ \\
5113557&F&2088&phabs*apec&$1.46\cdot10^{20}$&$0.60$&$6.40\cdot10^{50}$\\
5653243&K&316&phabs*apec&$\sim 0$&$0.80$&$5.86\cdot10^{52}$ \\
6761532&G&226&phabs*apec&$3.60\cdot10^{20}$&$0.63$&$2.43\cdot10^{52}$\\
7018708&G&2637&phabs*(apec+apec)&$6.80\cdot10^{20}$&$0.29,1.04$&$1.93\cdot10^{53},2.24\cdot10^{53}$ \\ 
8454353&M&1291&phabs*apec&$\sim 0$&$0.99$&$9.69\cdot10^{51}$ \\
8517303&K&1763&phabs*apec&$9.10\cdot10^{20}$&$1.06$&$1.10\cdot10^{54}$ \\
8518250&K&434&phabs*(apec+apec)&$\sim 0$&$0.94,0.23$&$1.24\cdot10^{51},1.88\cdot10^{51}$ \\
8520065&F&764&phabs*(apec+apec)&$2.33\cdot10^{20}$&$1.07,0.54$&$4.30\cdot10^{53},3.42 \cdot10^{53}$ \\ 
8584672&K&230&phabs*apec&$\sim 0$&$0.83$&$1.21\cdot10^{53}$ \\
8647865&F&433&phabs*apec&$\sim 0$&$0.49$&$1.60\cdot10^{52}$ \\
8713822&K&675&phabs*apec&$1.43\cdot10^{20}$&$1.05$&$2.58\cdot10^{54}$ \\
8842083&K&590&phabs*apec&$1.1\cdot10^{21}$&$0.76$&$7.01\cdot10^{50}$ \\
9048551&K&1634&phabs*apec&$\sim 0$&$1.07$&$8.91\cdot10^{51}$ \\
9048949&K&844&phabs*apec&$\sim 0$&$0.94$&$4.31\cdot10^{51}$ \\
9048976&K&285&phabs*apec&$\sim 0$&$0.96$&$4.11\cdot10^{51}$ \\
11971335&G&696&phabs*apec&$\sim 0$&$1.30$&$2.10\cdot10^{53}$ \\ \hline
\multicolumn{7}{l}{$^{(a)}$ {Spectral type has been evaluated from the $T_{\rm eff}$ according to \cite{2013ApJS..208....9P}.}} \\
\multicolumn{7}{l}{$^{(b)}$ {This is the sum of the counts in the three EPIC instruments.}} \\

\end{tabularx}
\end{center}
\end{table*}

\subsubsection{Considerations on the X-ray luminosity distribution}
\label{subsubsect:act_xray_bias} 

We compare the distribution of the X-ray luminosity in the range
$0.2-2.0\,{\rm keV}$ for the stars in our sample with those in the {\rm NEXXUS}
sample of \cite{2004A&A...417..651S}, which {\rm consists in a compilation} of coronal X-ray
emission for 
nearby 
late-type stars
based on the {\em ROSAT} observatory.
The distribution of the X-ray luminosity for our {main-sequence} sample and
for the one from \cite{2004A&A...417..651S} is plotted, for each spectral type and
for the whole sample, in Fig. \ref{fig:lx_histo}.

{To first order we consider NEXXUS as a volume-limited sample of nearby stars,} 
representing the full range of X-ray activity of the solar-like stellar population. 
{Note, however, that \cite{2013MNRAS.431.2063S} {showed} that even in as small a volume 
as $10$\,pc around the Sun about $40$\,\% of the M dwarfs have no X-ray detection in the RASS.} 
From the
comparison with our distribution, it is evident that our sample presents a
significant bias towards high X-ray luminosities 
most marked for the latest spectral types reflecting the mass (or $T_{\rm eff}$)
dependence of X-ray luminosity.   
The stars in our sample, which is flux-limited, are on average at a much greater
distance (see Fig. \ref{fig:lx_histo}, median distance for the whole sample: $\sim
\mathbf{500}\,{\rm pc}$). 
{\rm Therefore,} we interpret the bias {\rm of our sample towards active stars} 
as due to 
{\rm both the X-ray selection and the large distances.}
%
\begin{figure}
\begin{center}
\includegraphics[width=13.0cm,angle=270]{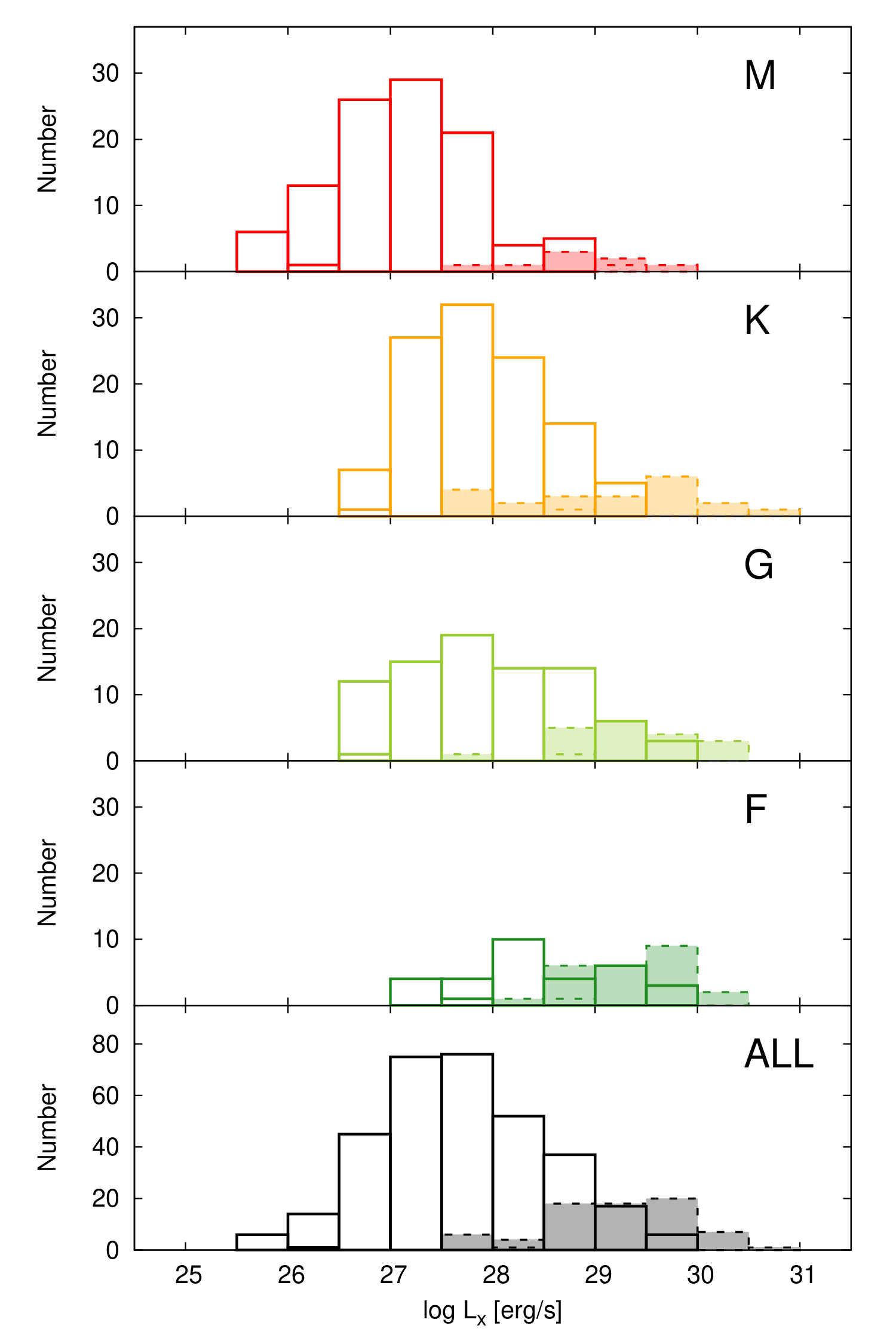}
\caption{Distribution of
the X-ray luminosity in the energy band $0.2-2.0\,${\rm keV}, {for the $102$ main-sequence stars} 
of the \kep - \xmm sample ({hatched histograms}) and for the 
{NEXXUS stars} from \cite{2004A&A...417..651S} (solid line) for each spectral type
{class separately and for all spectral types combined}.}
\label{fig:lx_histo} \end{center} \end{figure}

\subsubsection{X-ray light curves}\label{subsubsect:act_xray_lc} 

In order to study the variability of the X-ray
emission and to search for X-ray flares, we analyse the light curves provided 
by the EXTraS ({\em Exploring the X-ray Transient and
variable Sky}) project \footnote{The EXTraS project {\rm (www.extras-fp7.eu)}, aimed
at the thorough characterization of the variability of X-ray sources in archival {\em
XMM-Newton} data, {\rm was} funded within the EU seventh Framework Programme for a
data span of 3 years starting in January 2014. The EXTraS consortium is lead by INAF
(Italy) and includes other five institutes in Italy, Germany and the United Kingdom.}
\citep{2015arXiv150301497D}.  
{EXTraS was synchronized with 3XMM-DR4, while we are studying X-ray sources from 
3XMM-DR5 in the
{\em Kepler} field. Therefore, light curves for observations 0671230201 and 0671230601 
are not present in the public EXTraS database, but were produced for this work by the 
EXTraS team, using the EXTraS analysis pipeline.}

EXTraS provides
a set of uniform bin light curves
with different bin size, from $10\,${\rm s } up to $5000\,${\rm s} and an optimum bin
size (chosen to have at least $25$ counts in each bin), both for the source and the
background extraction regions.
{In addition,} EXTraS also provides light curves produced
via an adaptive binning, namely bayesian blocks  \citep{2013ApJ...764..167S}
algorithm for each source and for the related background region. This algorithm
starts with an initial set of cells defined on the basis of the number of events in
the source and background region, and provides a final set of different-duration
bins, each of which has a count rate that is not consistent, within $3\,\sigma$, with
the count rate of the adjacent bins.

In the EXTraS analysis pipeline, all
the source light curves, both the ones obtained with the uniform bin algorithm and
the bayesian blocks algorithm, are automatically fitted with a series of different
models that account for simple variability patterns (constant, linear, quadratic,
negative exponential, constant plus flare, constant plus eclipse). 
This is a standard algorithm that is part of the EXTraS pipeline, and it is applied to all
the light curves in the same way. Its purpose is to provide a first-step 
indication of the kinds of variability possibly present in the light curve, and not
to perform a {detailed} modeling of the variability features observed, 
nor to determine their parameters.
{All results are in the database, and online searches can be performed with the query 
form.}

\subsubsection{X-ray flaring} \label{subsubsect:act_xray_flares} 

For each X-ray  source in our
sample, we search for X-ray flares in the EXTraS light curves. First, we inspect 
the results of the fit performed on the light curves with
different variability models by the EXTraS pipeline. As stated above, these results
can be used for a first assessment for the kind of variability present in the light
curve. If at least one among the uniform bin or adaptive bin light curves is better
fitted by a constant plus flare model (higher null-hypothesis probability) than the other models, this is
a good indication of a possible flare. This flare model consists of a constant flux
level over which a simple flare profile is superimposed, that is a steep linear flux
increase, followed by an exponential decay.

The bayesian blocks light curves are particularly useful to detect flares, which
appear as one or more blocks that show a higher flux than the preceeding and
following blocks. If the flare occurred at the beginning or at the end of the
observation, the bayesian blocks light curve may show only the rise phase or the
decay phase. In this case, we rely on the uniform bin light curves to establish if
the event is a true flare or not. We also require that the flare was observed in at
least two of the EPIC instruments. The visual inspection of the light curves is
however crucial in order to recognize genuine flares, so we inspect all the available
EXTraS light curves for each star in our sample.

With this procedure, based on the EXTraS pipeline products we detect $6$ 
X-ray flares on $5$ stars, and by visual inspection we identify an additional
likely flare on a sixth star, KIC\,7018131 (Figs. \ref{fig:flare1},
and~\ref{fig:flare2}).  
In view of the low {count} statistics of the X-ray lightcurves some remarks on the
individual events are in order. KIC\,9048976 shows two possible flares: the bayesian
blocks algorithm shows one block with a higher flux at the beginning and at the end
of the light curve, 
impeding the observation of the full flare profile. KIC\,8909598 also shows an X-ray flare at
the end of the observation. The uniform bin light curves reveal    
a quite obvious flare profile both in PN and MOS2 cameras. However, this star does
not have a reliable main-sequence classification (unknown $\log{g}$), so we do not
consider it in the analysis. For KIC\,9048551 the bayesian block algorithm identifies one
flare event, but the uniform bin lightcurve shows evidence of substructure
contemporaneous with two events seen in the {\em Kepler} band.
%
%
\begin{figure} 
\begin{center}
\includegraphics[width=9.1cm]{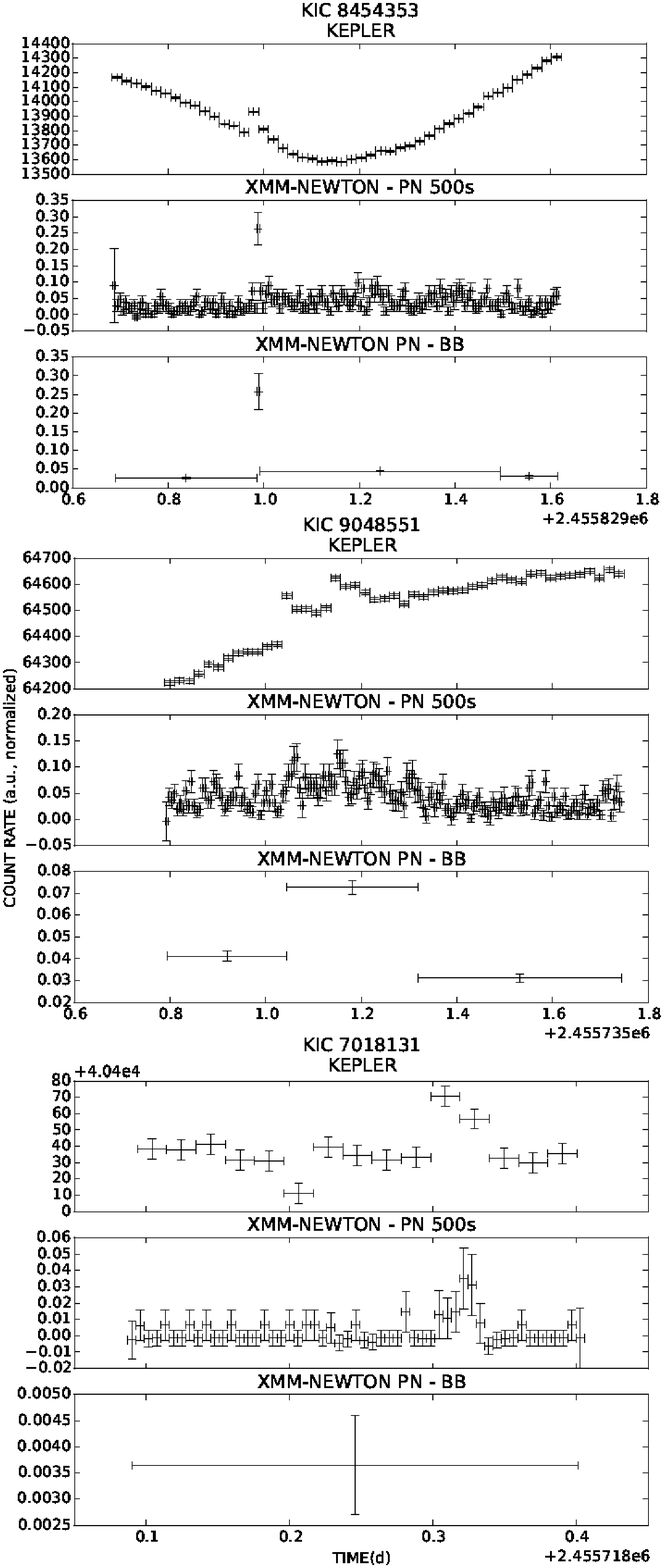} 
\caption{Simultaneous {\em Kepler} and {\em XMM-Newton} EPIC PN light curves for the 
stars showing a simultaneous X-ray and white-light flare.
The EPIC light curves
produced by the EXTraS pipeline with $500\,{\rm s}$ uniform binning and with the
bayesian blocks adaptive binning are plotted. 
{The flare in the \kep light curve of KIC\,7018131 corresponds to a bump in the 
EPIC uniform bin lightcurve which is, however,} 
not significant at a $3\sigma$ level over the baseline, and not detected
with the bayesian block algorithm.
}
\label{fig:flare1} 
\end{center} 
\end{figure} 

\begin{figure} 
\begin{center}
\includegraphics[width=9.3cm]{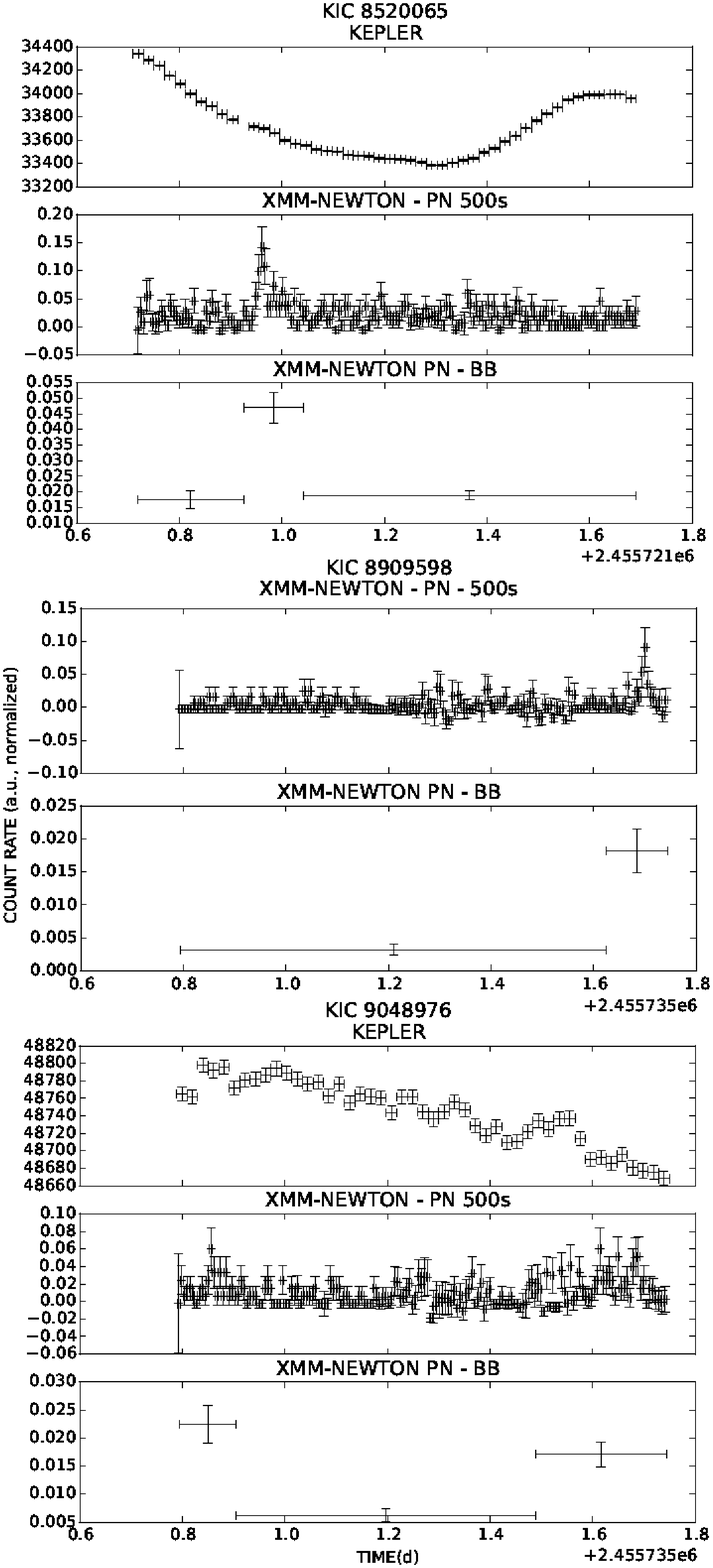} \caption{{\em XMM-Newton} EPIC PN
light curves for the X-ray flares without a white-light counterpart detected by our
flare-detection algorithm. Simultaneous {\em Kepler} light curves are not available
for KIC\,8909598.
}
\label{fig:flare2} 
\end{center} 
\end{figure} 

The X-ray flares shown in Fig.~\ref{fig:flare1} have simultanous white-light
flares in the {\em Kepler} light curves (see Sect. \ref{subsect:results_kepler}). 
{\rm Fig.~\ref{fig:flare2} displays the light curves for the X-ray flares without a
simultaneous white-light counterpart revealed with the flare detection algorithm
described in Sect.~\ref{subsect:kepler_spot}.  
Evidence for a weak increase of the
optical brightness at times of the X-ray flare is, however, seen in all the
contemporaneous {\em Kepler} lightcurves.} 

In Table \ref{tab:x_flares} we report
the start time of the X-ray flares (col.~2), according to the bayesian blocks light
curve except for KIC\,7018131, for which we infer the start time from the $500\,{\rm
s}$ uniform bin light curve of {\rm EPIC/pn}, since the bayesian blocks light curve
has only one block. We also report the quiescent (col.~3) and peak (col.~4) X-ray
count rate, taken from the $500\,{\rm s}$ uniform bin light curve of {\rm EPIC/pn},
together with the overall number of white-light flares observed in the whole {\em
Kepler} light curve for the star (col.~5), its white-light flare frequency (col.~6),
the average, maximum and minimum peak amplitude of the \kep flares ($A_{\rm peak}$,
photometric {ratio} between the peak bin and the baseline of the flare, cols.~7 and~8). 
{This parameter is discussed in more detail in Sect.~\ref{subsubsect:results_kepler_flares}.} 

\begin{table*} 
\begin{center} 
\tiny{\caption{Parameters of the X-ray flares and
{\em Kepler} flare characteristic of the X-ray flaring stars. The start time of each
X-ray flare is reported, together with the peak and off-flare count rate. The number
and the {frequency} of the white-light flares observed for the star is also given,
together with the average, minimum and maximum peak amplitude of all its white-light
flares in the {\em Kepler} light curve.} 
\label{tab:x_flares}
\begin{tabularx}{\textwidth}{lcccccccccr} \hline 
KIC\_ID&Start time&$Rate_{\rm
X,quiesc}$&$Rate_{\rm X,peak}$&$N_{\rm opt,flares}$&$F_{\rm opt,flares}$&$<\log{A_{\rm peak}}>$
& Min-max $\log{A_{\rm peak}}$\\ &(Julian day)&{\rm (cts/s)}&{\rm
(cts/s)}&&{\rm ($d^{-1}$)}&&\\ \hline 
7018131&2455718.31&0.006$\pm$0.011&0.035$\pm$0.019&5&0.005 & -2.79 & -2.89 / -2.67 \\
8454353&2455829.96&0.036$\pm$0.020&0.263$\pm$0.048&297&0.211 & -1.86 & -0.36 /-2.69 \\
8520065&2455721.96&0.020$\pm$0.016&0.140$\pm$0.040&0&0&0&0\\
8909598&2455736.55&0.004$\pm$0.011&0.090$\pm$0.003&7&0.022 & -1.83 & -1.43/-2.21 \\
9048551&2455736.03&0.033$\pm$0.016&0.125$\pm$0.028&233&0.164 & -2.41 & -1.47/-2.92 \\
9048976&2455735.94,2455736.80&0.009$\pm$0.013&0.06$\pm$0.02,0.060$\pm$0.024&137&0.097
& -2.37 & -1.57/-2.91 \\ 
\hline 
\end{tabularx}
}
\end{center}
\end{table*}

\subsection{UV activity}\label{subsect:act_uv}

The KIC provides
UV magnitudes obtained with the {Galaxy Evolution Explorer (GALEX)}. 
The GALEX satellite performed imaging in two UV bands, far-UV (henceforth FUV; 
$\lambda_{\rm eff} =1516 \,${\rm \AA{}}, $\Delta\lambda=268\,${\rm \AA{}} , and
near-UV (henceforth NUV; $\lambda_{\rm eff} = 2267\,${\rm \AA{}},
$\Delta\lambda=732\,${\rm \AA{}}). {The KIC gives a NUV detection for $71$
stars from our main-sequence sample (i.e. $70$\,\%) and a FUV detection for $20$
main-sequence stars ($20$\,\%).} 
All but one of the stars with a FUV detection also have a NUV detection.
{The individual values for the observed NUV and FUV luminosities 
are listed in cols.~3 and~4 of Table~\ref{tab:xray}.}

The SED fit provides UV fluxes for the stellar photosphere. In order to validate
these measurements, we convert the NUV de-absorbed photospheric fluxes obtained
from VOSA into absolute magnitudes and compare the relation between these magnitudes
and the $B-V$ colour (derived from \citealt{2013ApJS..208....9P}) with the
analogous relation observed for a set of photospheric NUV magnitudes provided by
\cite{2011AJ....142...23F} in their Table~1. Fig.~\ref{fig:findeisen} shows that
our values are in good agreement with the ones of \cite{2011AJ....142...23F}.

For some stars in our sample the observed GALEX NUV and FUV fluxes are
significantly higher than the prediction of the best-fitting photosphere model. This
UV excess, i. e. the positive difference between the observed UV flux ($f_{\rm
UV,obs}$) and the photospheric flux of the BT-Settl model in the same UV band
($f_{\rm UV,ph}$) represents the UV emission associated with magnetic
activity processes in the stellar chromosphere. Following \cite{2013MNRAS.431.2063S},
we calculate the corresponding UV activity index as 
\begin{equation}
AI'_{\rm UV}=\frac{f_{\rm UV,exc}}{f_{\rm bol}}=\frac{f_{\rm UV,obs}-f_{\rm UV,ph}}{f_{\rm bol}} 
\label{eq:exc}
\end{equation} 
where $f_{\rm UV,exc}$ is the UV excess flux
attributed to activity, and $f_{\rm bol}$ is the bolometric flux. The bolometric flux
is {obtained from} interpolation on the DSEP isochrones, as described in Sect.~\ref{sect:params}.

We calculate the UV excess in both the GALEX FUV and NUV
band, where available. After trying different values and visually inspecting the SED,
we find that a threshold of $13\%$ on the ratio between the excess flux and the
photospheric expected flux is the best choice to select the stars with a true UV
excess. {We found $45$ main-sequence stars with a NUV excess, ten of them
display also a FUV excess, and an additional $4$ have a FUV excess but no NUV excess.} 
{The NUV and FUV excess luminosities are provided in cols.~5 and~6 of Table~\ref{tab:xray}}.

\begin{figure}
\begin{center}
\includegraphics[width=8.5cm]{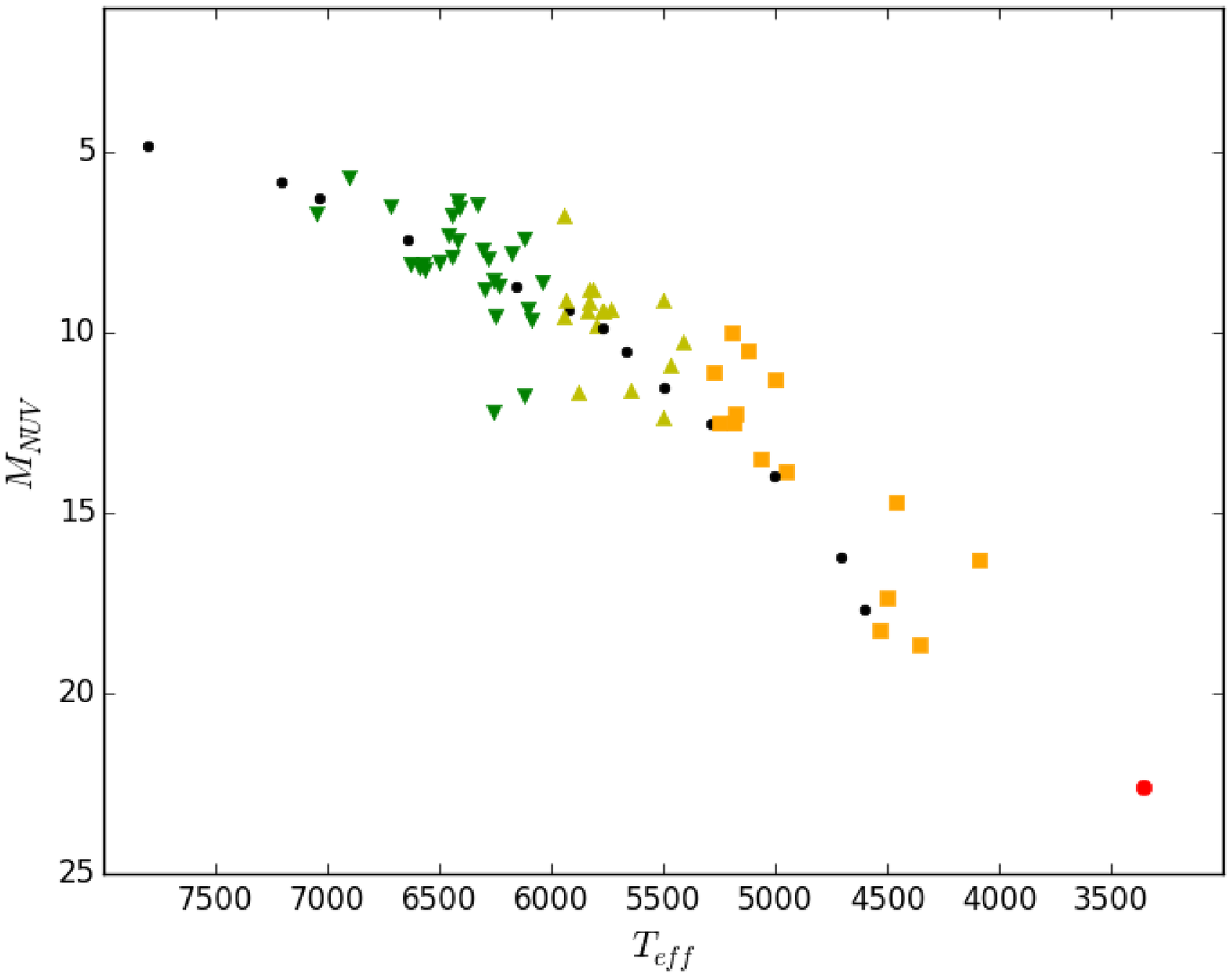}
\caption{Absolute NUV magnitude versus Johnson colour {\em B-V} for the stars in our
sample with a GALEX NUV detection, for each spectral type, and for the values of
Table 1 in \cite{2011AJ....142...23F} (black dots). Different symbols represent the
spectral type; red circles: M, orange squares: K, light-green triangle up: G, dark-green
triangle down: F.}
\label{fig:findeisen} 
\end{center} 
\end{figure}

%
%
%
%
%
%

\section{Results and discussion} \label{sect:results}

\subsection{Rotation periods}\label{subsect:results_prot} 

For $74$ main-sequence stars out of $102$ ($73\%$) the
{\em Kepler} light curve is dominated by the rotational brightness modulation due to
starspots, with a period in the explored range $P_{\rm rot}<90\,${\rm d}. For the
individual spectral types, the fraction of {main-sequence} stars with a rotational brightness
modulation is: 
{A $66$\,\% (2/3), F $65$\,\% (24/37), G $73$\,\% (19/26), K $75$\,\% (21/28) 
and M $100$\,\% (8/8).}
Our period detection
rate is higher with respect to other studies based on {\em Kepler} data
(\citealt{2013ApJ...775L..11M,2014ApJS..211...24M}, $37\%$ in the range $3500\,{\rm
K}<T_{\rm eff}<6500\,{\rm K}$, F $\sim 27\%$, G $\sim 25\%$, K $\sim60\%$, M
$\sim80\%$; \citealt{2016MNRAS.tmp.1060S}, $73\%$ for M {dwarfs}). This is probably an
effect of the selection bias towards active (and thus strongly spotted) stars 
{as well as the removal of giant stars from our sample.}

In Table~\ref{table:var} a summary of the number of stars with
rotation period from our analysis is reported for every spectral type, together with
the number of stars with {other types of} brightness modulation patterns. 
{We refer to 
Table~\ref{table:params}
for the classification of the photometric variability for each star
in the sample.} 

Only $54$ stars out of $74$ have
previously reported periods from \cite{2014ApJS..211...24M}, obtained from the same
{\em Kepler} light curves. For the $54$ stars for which a rotation period is reported in
\cite{2014ApJS..211...24M}, 
those periods are consistent with our values within uncertainties. The
remaining $22$ stars do not have any previously determined period in the
literature.

%
%
%

\subsection{Activity-rotation relation} \label{subsect:results_actrot}

In Fig. \ref{fig:lx_prot}, we present the relation between the X-ray luminosity in
the soft energy band $0.2-2.0\,${\rm keV} and the rotation period, together with the
relation between the X-ray activity index ({defined in} Eq. 3) and the Rossby number. 
A clear decrease in X-ray activity levels is observed for slower rotators.
This effect is more evident in the $AI_{\rm x}$ versus Rossby number plot, and
produces a `kink' in the distribution, which can be seen quite clearly in the overall
sample of all stars, and also in the subsamples of K stars, while for M, G and F
stars it is not {obvious as these subsamples comprise only a limited range of
rotation rates}. The `kink' suggests the presence of a correlated regime for
slow rotators, and of a saturated regime for fast rotators, with a separation
occurring at $\sim 8\,{\rm d}$ (from visual inspection). The wide range of $L_X$ and 
$AI_{\rm X}$
independent of $P_{\rm rot}$ for the (generally fast rotating) F stars is remarkable:
it is possible that the F stars present a decoupling between rotation and activity
because of their shallow convective zones. This group may also include active binary
stars (RS\,CVn) with unknown contribution to the X-ray emission from the cool
companions.

In Fig.~\ref{fig:lx_prot} we show for comparison the literature
compilation from \cite{2011ApJ...743...48W} {\rm and} the {previous} 
empirical relations obtained by \cite{2003A&A...397..147P} {based on a small sample
with mostly spectroscopic rotation measurements} (solid lines). 
The rotation periods 
in \cite{2011ApJ...743...48W} {\rm were} collected from several works which use both
spectroscopic and photometric techniques, {\rm and the} X-ray fluxes {\rm were} 
obtained from the analysis of data taken from different missions, such as 
{\em XMM-Newton} and {\em ROSAT}. We find excellent agreement of our {\rm more
homogeneous} data with that study. 
{\rm Especially, the scatter clearly decreases when X-ray luminosity and rotation period 
are replaced by $AI_{\rm x}$ and $R_0$, respectively.}
{Contrary to previous work where no uncertainties were estimated, 
we present here conservative error bars for our sample.
These are dominated by the uncertainties in the distances 
of stars without Gaia parallax that are derived from mapping the 
stars in the $\log{g} - \log{T_{\rm eff}}$ diagram onto the Dartmouth isochrones 
(see Sect.~\ref{subsect:params_dsep}), i.e. they ultimately go back to the uncertainties 
in the spectroscopic parameters.}

%
\begin{figure*} 
\begin{center}
\includegraphics[width=22.8cm,angle=270]{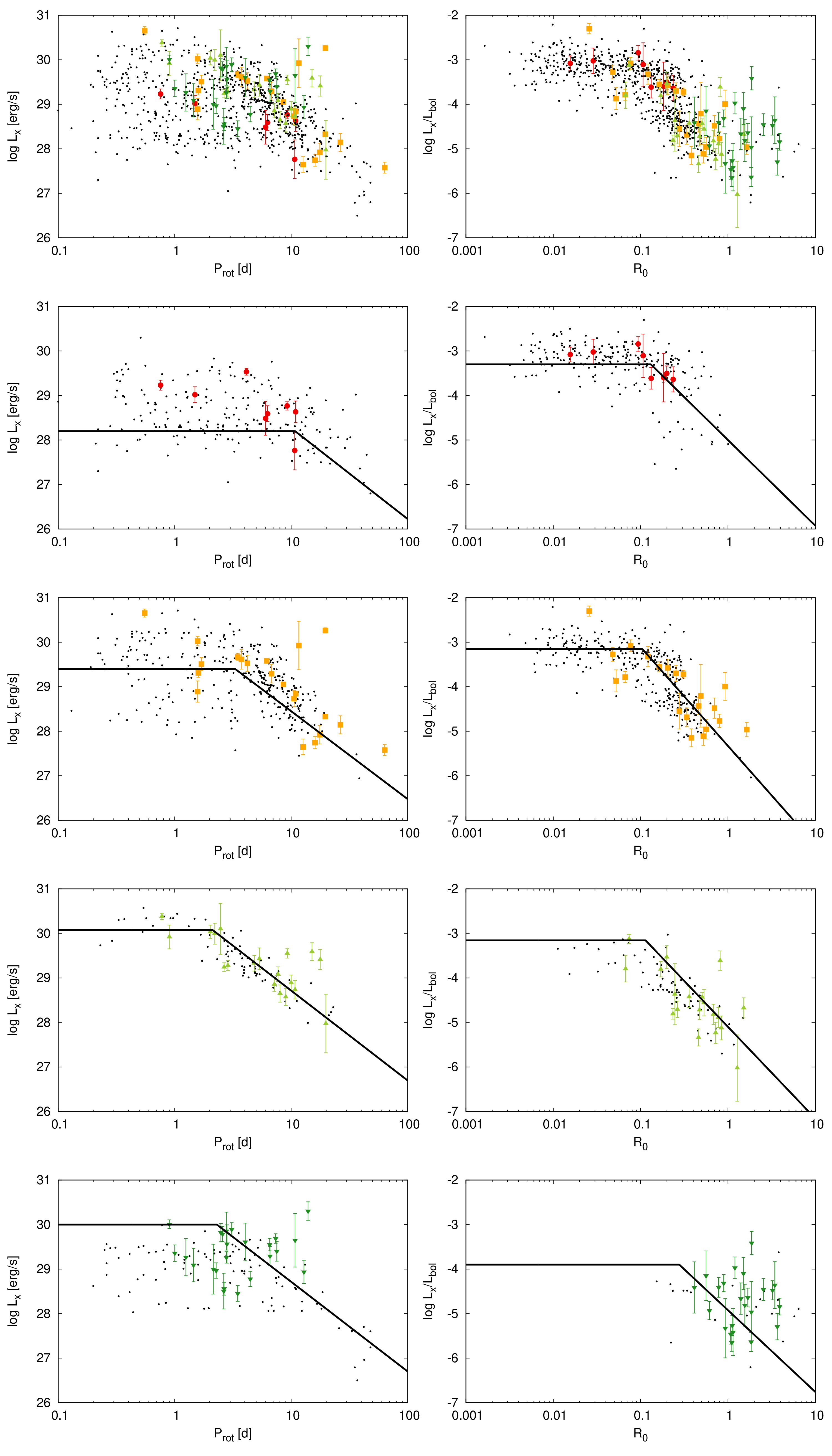}
\caption{{\rm X-ray activity versus rotation for the full sample, M-, K-, G-, and F-type
stars (from top to bottom).} 
Left panel: X-ray luminosity in the energy band $0.2-2.0\,${\rm keV} versus \kep 
Right panel: X-ray activity index versus Rossby number. 
Colours and symbols 
{\rm follow the convention defined} in Fig. \ref{fig:findeisen}. 
The stars in the sample of \cite{2011ApJ...743...48W} are
shown as small black dots. The solid lines represent the best-fit relations between
X-ray emission and rotation period found by \cite{2003A&A...397..147P} for 
stellar mass ranges {\rm corresponding approximately to spectral types}.}  
\label{fig:lx_prot} 
\end{center} 
\end{figure*} 

{The rotation-activity relation of M dwarfs in our sample can be compared to that
of M dwarfs observed in the K2 mission studied by \cite{2016MNRAS.tmp.1060S} with an 
analogous approach. In Fig.~\ref{fig:K2_ours_prot} we show the $L_{\rm x} - P_{\rm rot}$
relation for the two samples. 
Our X-ray selection evidently excludes slowly rotating M stars in the non-saturated regime. 
This can also be seen from the comparison with the 
bimodal relation suggested by \cite{2003A&A...397..147P} (solid black line 
in Fig. \ref{fig:lx_prot}) which is, however, itself extremely poorly defined. Recent
M dwarf studies by \cite{2016Natur.535..526W} and \cite{2018MNRAS.479.2351W}, also shown
in the figure, used 
rotation periods measured with ground-based instruments. They 
cover the long-period `correlated' region and are complementary to our {\em Kepler} study.}  
\begin{figure} 
\begin{center}
\includegraphics[width=6.5cm,angle=270]{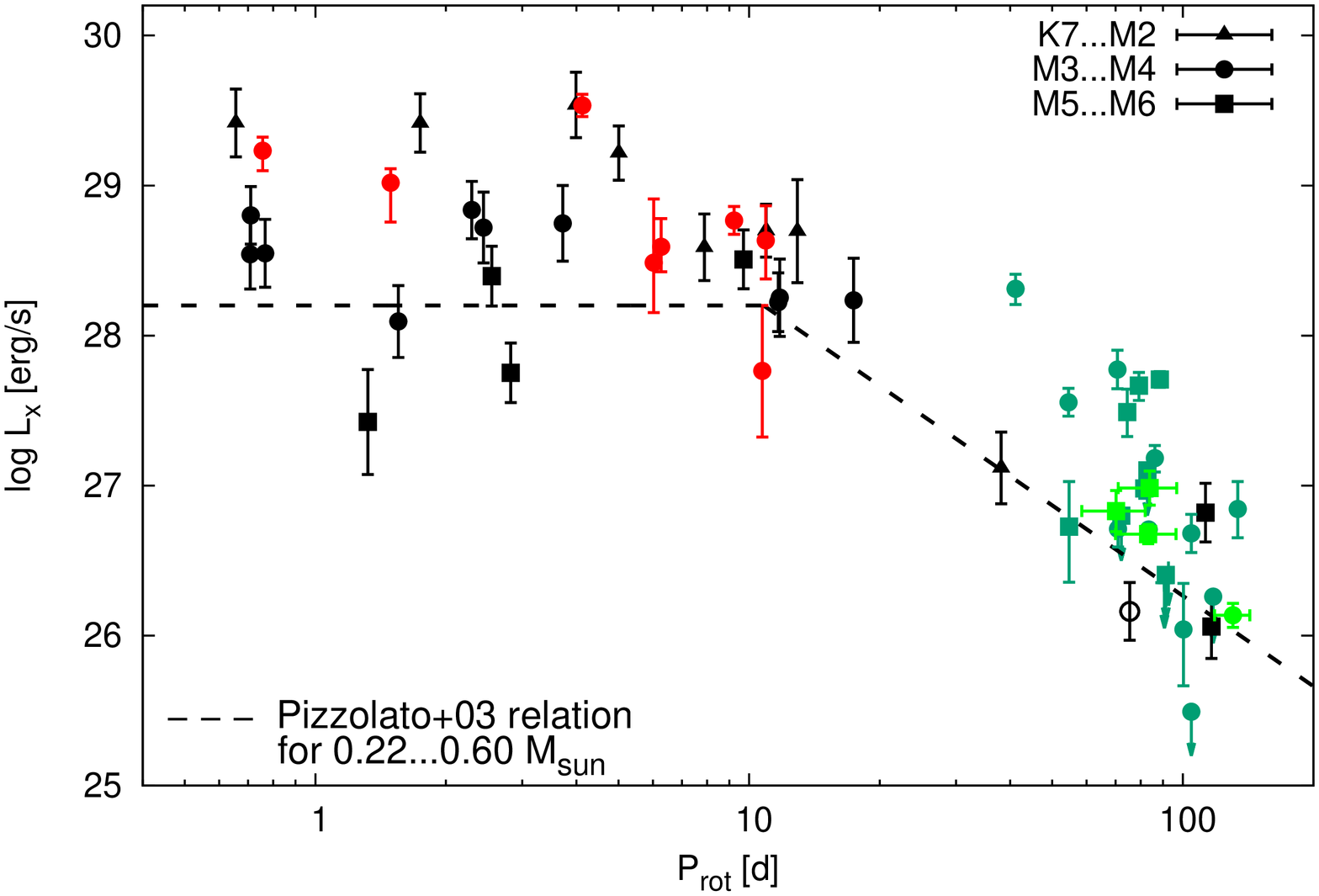}
\caption{$L_{\rm X}$ vs $P_{\rm rot}$ for M dwarfs: {red -- this work, black --  
K2 sample from \citet{2016MNRAS.tmp.1060S} (their Fig.~15), 
green -- stars published by \citet{2016Natur.535..526W} and \citet{2018MNRAS.479.2351W}}.  
The solid line represents the best-fit rotation/activity relation 
from \cite{2003A&A...397..147P} in the range $0.22-0.60\,M_{\odot}$.} 
\label{fig:K2_ours_prot} 
\end{center} 
\end{figure}

The scatter observed in {the saturated regime of}
$L_{\rm X}$ and $AI_{\rm x}$ for M stars {\rm of given $P_{\rm rot}$} is 
{\rm large} but consistent with
that observed by \cite{2011ApJ...743...48W}, \cite{2003A&A...397..147P} and 
\cite{2016MNRAS.tmp.1060S}, and probably due, at least in part, to the spectral type
distribution {within the M class}, with cooler stars having lower X-ray luminosity.
The scatter decreases when considering the relation $AI_{\rm x}$ versus $R_0$.
Because of the relatively low statistics in our sample, in particular the low number of stars in
the correlated regime, it would be difficult to establish with good confidence the
turnover point between the correlated and saturated regime, nor the slope of the
correlated part of the relation.


\subsection{{\em Kepler} activity diagnostics} \label{subsect:results_kepler}

\subsubsection{Optical flares}\label{subsubsect:results_kepler_flares}

{We explore here the optical flaring activity measured in the {\em Kepler} lightcurves
of the spotted stars.
This analysis must consider that the sensitivity for detecting flares depends on the
(quiescent) brightness of the star and is different for each star. 
The criteria in our definition of flare include a $\geq 3\,\sigma$ upward deviation 
from the `flattened' lightcurve (see Sect.~\ref{subsect:kepler_spot}). As a result of the
relation between $K_{\rm p}$ and spectral type in our sample (evident in 
Fig.~\ref{fig:sflat_kp}), the minimum measurable 
flare amplitude ($A_{\rm peak}$) -- defined here as the relative brightness difference 
between the flare peak and the `flattened' lightcurve -- shows a trend with spectral type 
(Fig.~\ref{fig:peakampl_sflat}).}\\
\begin{figure} 
\begin{center}
\includegraphics[width=7.0cm, angle=270]{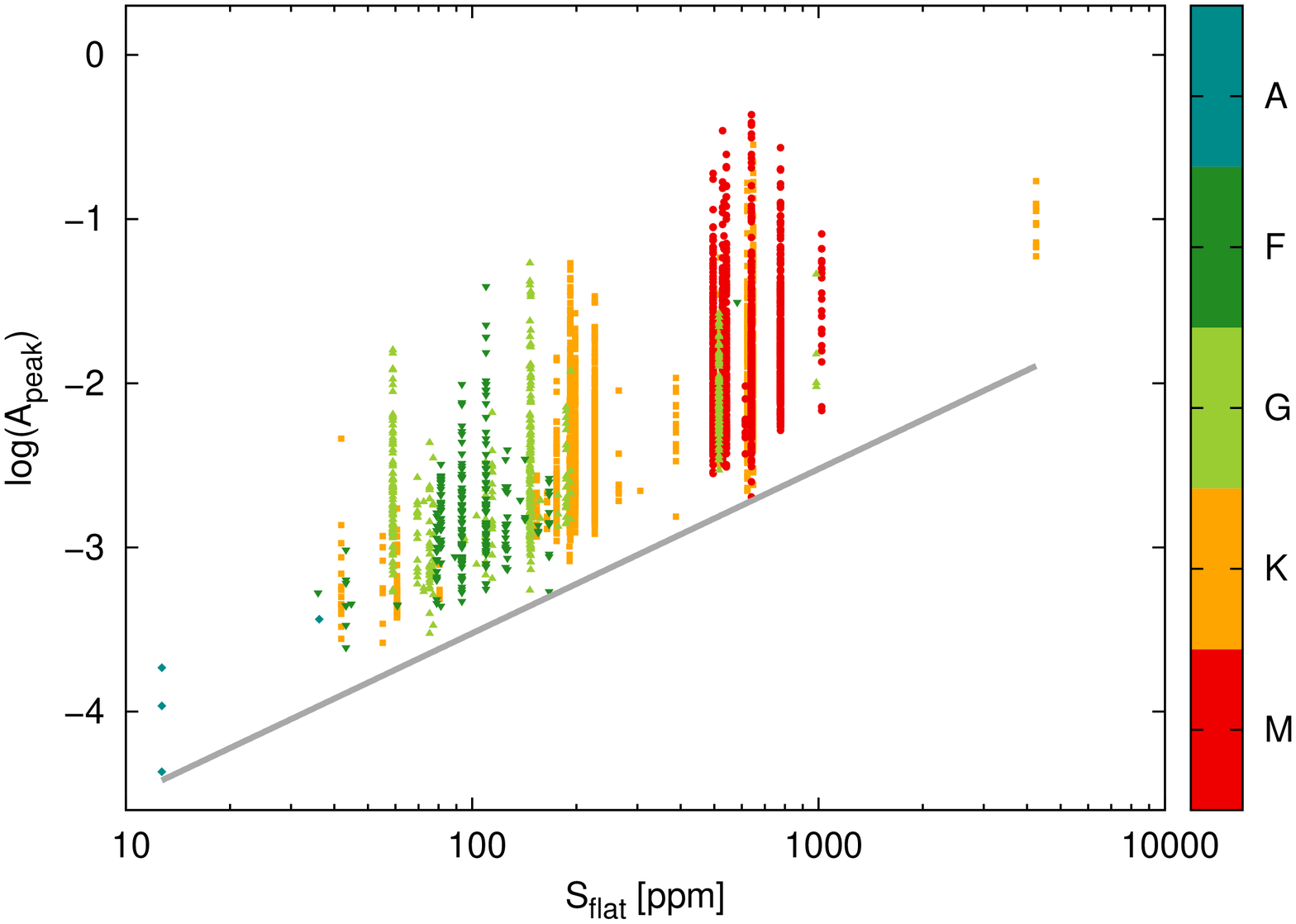} 
\caption{{Relative peak amplitudes of all detected optical flares in the {\em Kepler} lightcurves
measured with respect to the `flattened' lightcurve. The grey line denotes our threshold
for flare detection set to $3 \cdot S_{\rm flat}$. 
}} 
\label{fig:peakampl_sflat} 
\end{center} 
\end{figure}
\indent {For a meaningful comparison of our sample which covers a large range in brightness
we must convert relative quantities, such as $A_{\rm peak}$ and $S_{\rm flat}$, to 
absolute ones. The {\em Kepler} photometry is not flux calibrated. However, 
an approximate brightness can be associated  to the `flattened' lightcurve assuming that 
this normalized `quiescent' emission corresponds to the {\em Kepler} magnitude of the star. 
We convert $K_{\rm p}$ to flux using the zero-point and effective {bandwidth} 
provided at the filter profile service of the 
{\em Spanish Virtual Observatory (SVO)}\footnote{http://svo2.cab.inta-csic.es/svo/theory/fps/}. 
Then we apply the distances derived in 
Sect.~\ref{subsect:params_dsep} to obtain the `quiescent' luminosity in the {\em Kepler} band, 
$L_{\rm K_p,0}$. 
Similarly, the flare amplitude is converted from its relative value ($A_{\rm peak}$) to
a luminosity, $\Delta L_{\rm F,K_p} = L_{\rm K_p,0} \cdot A_{\rm peak}$. 
The resulting relation between flare amplitude (in erg/s) and `quiescent' luminosity
is shown in Fig.~\ref{fig:Lampl_L0}. The lower envelope is defined by our flare
detection threshold which is marked as a horizontal bar for each star and which is
rising with increasing stellar luminosity. 
The A-type stars form an exception to this trend. They show lower flare amplitudes than expected
for their $L_{\rm K_p,0}$. This finding can easily be explained when the observed flares
are attributed to unknown and unresolved later-type companion stars. In this scenario
the actual $L_{\rm K_p,0}$ value of the flare-host would be lower making the true amplitude
$\Delta L_{\rm F,K_p}$ higher than measured. This would shift the data points to the
left and upwards in Fig.~\ref{fig:Lampl_L0}.} 
\begin{figure} 
\begin{center}
\includegraphics[width=7.0cm, angle=270]{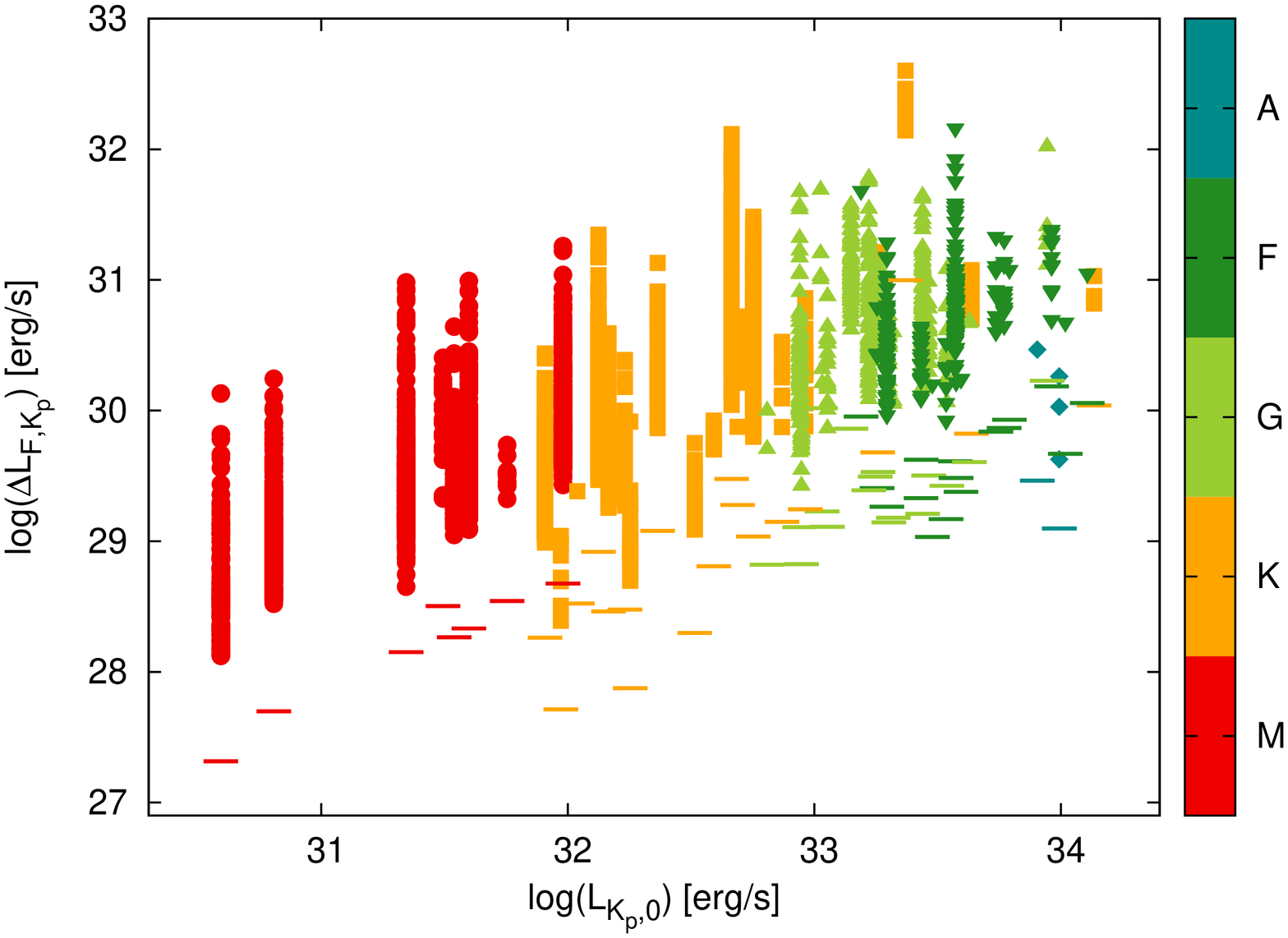} 
\caption{{Absolute flare amplitudes versus quiescent luminosity represented by the
{\em Kepler} magnitude associated with the `flattened' lightcurve; see text for details.}} 
\label{fig:Lampl_L0} 
\end{center} 
\end{figure}

{For a given star, the range of flare luminosities, $\Delta L_{\rm F,K_p}$, 
extends up to $\sim 2$ orders of magnitude 
above the amplitude of the minimum observable flare. The range of flare amplitudes is
smaller for G and F stars but this is probably related to the lack of sensitivity for
the detection of low-luminosity flares. This is evident from consideration of the
$S_{\rm flat}$ values 
which set the detection threshold (see Fig.~\ref{fig:peakampl_sflat} and~\ref{fig:Lampl_L0}).} 

{One of the aims of our study is the investigation of a connection between 
flaring activity and rotation rate. In their analogous study on M dwarfs observed in the 
K2 mission, \cite{2016MNRAS.tmp.1060S} have found a sharp transition in the optical 
flaring behavior at a period of $\sim 10$\,d. 
This transition 
is difficult to probe 
with this X-ray selected {\em Kepler} sample for two reasons: (1) the small number of
slow rotators and (2) the broad range of $K_{\rm p}$ translating into a dishomogeneous
flare detection threshold. 
If we restrict our sample to M dwarfs which span a relatively narrow range in 
optical brightness (c.f. Fig.~\ref{fig:sflat_kp}), the relative flare amplitudes and
the flare frequencies are
consistent with the results obtained by \cite{2016MNRAS.tmp.1060S} but we are covering
only the fast periods up to the presumed transition (see Fig.~\ref{fig:flares_prot_mstars}). 
The average flare frequency of the M
stars in our sample ($0.12\,{\rm flares/d}$) 
is in excellent agreement with the average
flare frequency in \cite{2016MNRAS.tmp.1060S} calculated over the $P_{\rm rot}$ range
($P_{\rm rot}\lesssim10\,{\rm d}$) covered by the {\it Kepler} M stars ($0.11\,{\rm flares/d}$).
This shows that our {\it Kepler} sample, albeit highly incomplete at the slow rotation / 
low activity side, is representative for fast rotating and \underline{active} M dwarfs.} 
%
\begin{figure} 
\begin{center}
\includegraphics[width=6.0cm, angle=270]{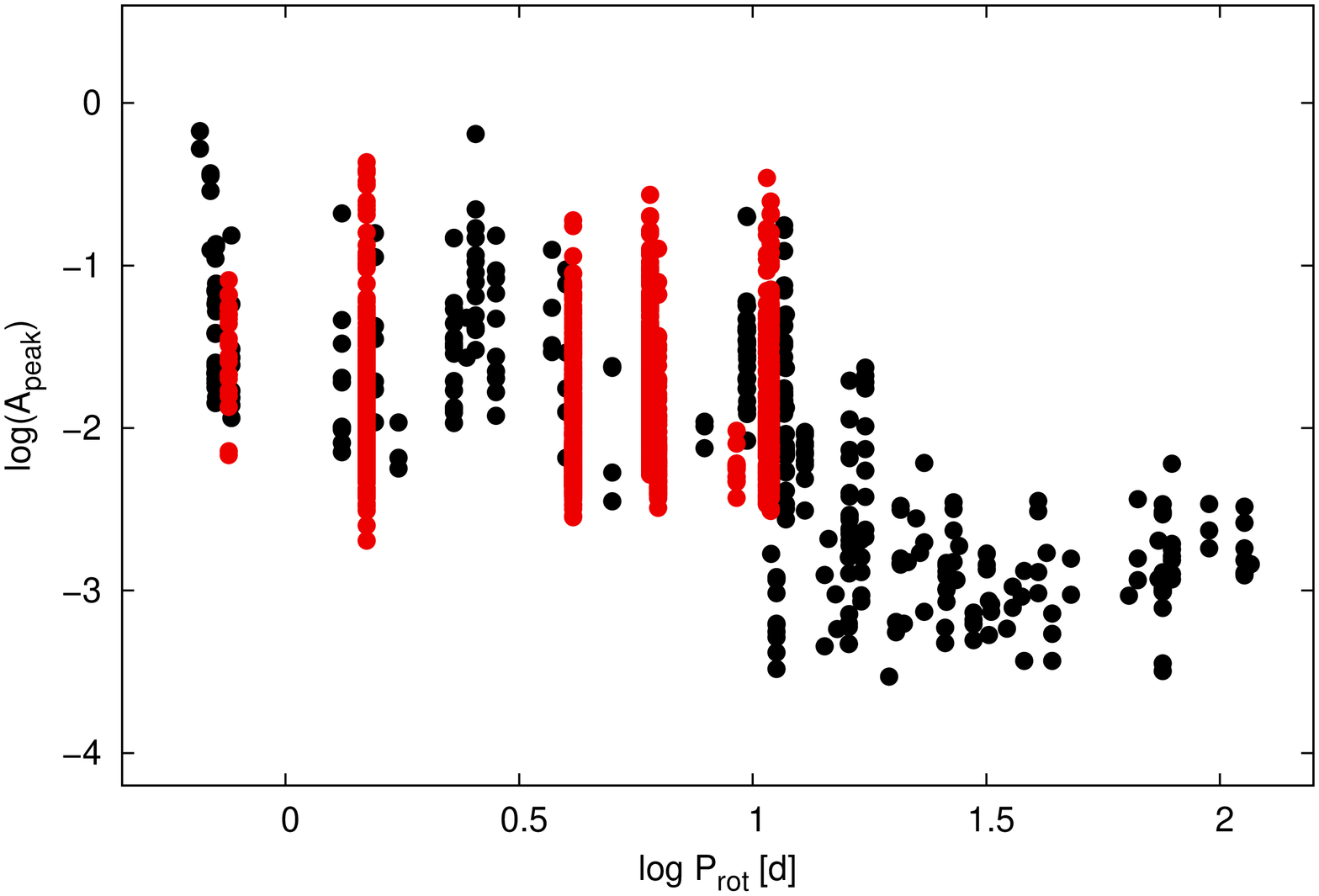}
\includegraphics[width=6.2cm, angle=270]{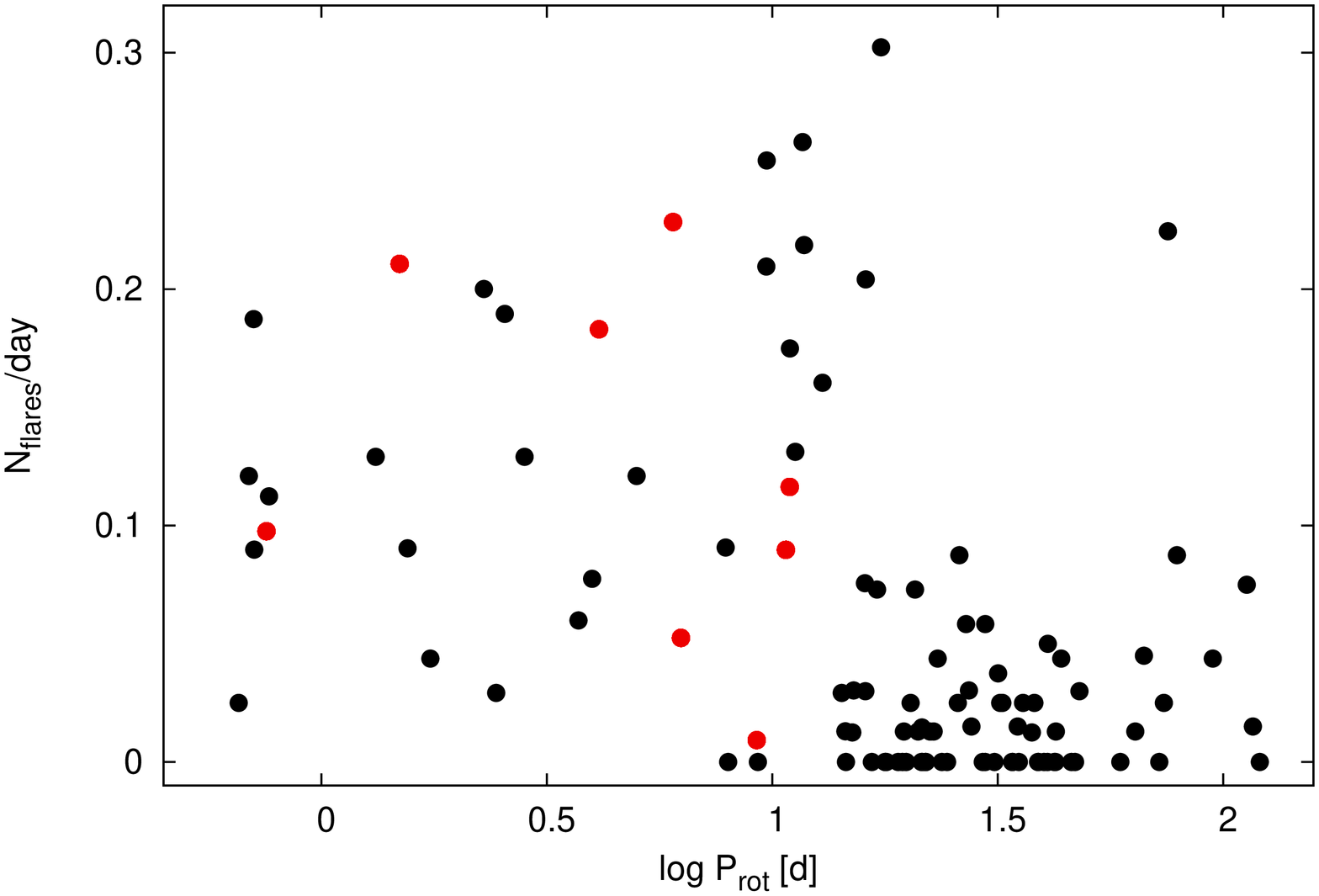}
\caption{{Relative flare amplitudes ($A_{\rm peak}$; top panel) and flare frequency (bottom panel)
versus rotation period for the M stars in our {\em Kepler} sample (red) and for the 
M stars of the K2 sample presented by \cite{2016MNRAS.tmp.1060S} (black).}} 
\label{fig:flares_prot_mstars} 
\end{center} 
\end{figure}

Analogous to the study of \cite{2016MNRAS.tmp.1060S} on K2 lightcurves, the
sampling time of $29.4\,$min of the {\em Kepler} lightcurves used in this work, 
together with the characteristics of the flare search algorithm 
(see Sect.~\ref{subsect:kepler_rot}), 
prevent the detection of flares with duration below 
$\sim 1\,{\rm h}$. As a consequence, we detect mostly flares with large-amplitude and relatively
long duration. It can be suspected that there is a significant population of smaller
and shorter flares that cannot be observed, due to the long cadence of the light
curves and to the flare-extraction pipeline. Therefore, the flare frequencies of 
Fig.~\ref{fig:flares_prot_mstars} 
{\rm and Table \ref{tab:mean_of_photact}} 
likely represent a lower limit to the actual values.
\begin{table}
\begin{center}
\caption{{\rm Mean values and standard deviations
measured for photometric activity diagnostics in the \kep / \xmm sample.}} 
\label{tab:mean_of_photact}
\begin{tabular}{llll} \hline
SpT & $N_{\rm f} /day$ & $\log{R_{\rm per}}$\,[\%] & $S_{\rm ph}$\,[ppm] \\ \hline
A & 0.001 & $-1.02$ & $3.47\cdot10^2$ \\
F & 0.01 & $-0.74$ & $1.10\cdot10^3$ \\
G & 0.02 & $0.01$ & $5.39\cdot10^3$ \\
K & 0.04 & $0.25$ & $8.69\cdot10^3$ \\
M & 0.12 & $0.46$ & $1.10\cdot10^4$ \\ \hline
\end{tabular}
\end{center}
\end{table}

\subsubsection{X-ray versus white-light flares}\label{subsubsect:results_kepler_xray} 

As described in Sect. \ref{subsubsect:act_xray_flares} 
we find $7$ X-ray flares on $6$ stars. For these events the count rate at the
flare peak (as measured from the EXTraS bayesian light curves) is a factor $3-7$ higher
than during the non-flaring, quiescent time-intervals. As a result of the large
distances of the {\em Kepler} stars, the X-ray count rates are small and the
low counts statistics prevent a detailed quantitative analysis such as the
determination of decay time and total flare energy. 

All the stars with an X-ray flare except KIC\,$8909598$ are classified as main-sequence: 
one  M-type, three
K-type and one F-type star. We do not consider KIC\,$8909598$ in the further analysis. All 
but the F star (KIC\,$8520065$), show a clear rotational modulation in their
{\em Kepler} light curves. This latter one is according to our analysis probably also
a rotator, but the variability pattern is unclear, and it is not possible to estimate
a reliable rotation period.

We compare the X-ray flaring with the
white-light flaring activity. Of the four stars with a detected rotation
period, three show a large number of white-light flares in the {\em Kepler}
lightcurves, and are among the stars with the highest optical flare frequency.  
This underlines the strong connection between X-ray and
optical flaring mechanism:  
Since the X-ray observations are relatively short in duration, 
X-ray flares are observed predominantly in the light curves of the stars which
show a higher optical flare frequency.
%
There are simultaneous {\em XMM-Newton}/{\em Kepler} observations for $69$
stars in our sample. All the X-ray flaring stars in our sample (except for KIC\,$8909598$, 
which, as mentioned, we exclude from the analysis) have a \kep observation simultaneous 
with the \xmm observation in which the X-ray flare was detected.  
No white-light flares without X-ray counterpart were {\rm detected} during the
simultaneous X-ray/white-light observations.\\
\indent Our automated \kep flare-detection algorithm finds
$3$ events occurring within $\sim 1\,{\rm h}$ from the observed X-ray flares  
(see Fig. \ref{fig:flare1}).
However, visual inspection shows that the other X-ray flares likely have optical 
counterparts as well, which
have remained below the detection threshold of our automatic algorithm. 
Finally, a simultaneous event, detected only by visual inspection of the X-ray
(and {\em Kepler}) lightcurves, occurred on one star. 
To summarize, of the {$6$} X-ray flares observed 
from rotating dwarf stars, there is
evidence for a contemporaneous optical event in all cases, but only for three of them
is the associated white-light flare clearly detected (one M-type and two
K-type stars). 



\subsubsection{Other \kep activity diagnostics}\label{subsubsect:results_kepler_other}

{Here we examine the distribution of the photospheric activity diagnostics 
$S_{\rm ph}$ and  
$R_{\rm per}$ (see Sect.~\ref{subsect:kepler_activity} for their definition).  
As mentioned above, as a result of the bias to active stars, our {\em Kepler} sample  
does not allow us to study the transition of the optical activity diagnostics 
at $P_{\rm rot} \sim 10$\,d revealed in the K2 data of nearby M dwarfs. 
Fig.~\ref{fig:sph_K1_K2} shows our M dwarf sample compared to the one studied by
\cite{2016MNRAS.tmp.1060S}. Similar to the results on optical flares, the rotation cycle
amplitude ($R_{\rm per}$) and the overall variability of the lightcurve ($S_{\rm ph}$)
of the two samples follow the same distribution in the fast-rotator regime.\\ 
\begin{figure}[!htb] 
\begin{center}
\includegraphics[width=6.5cm,angle=270]{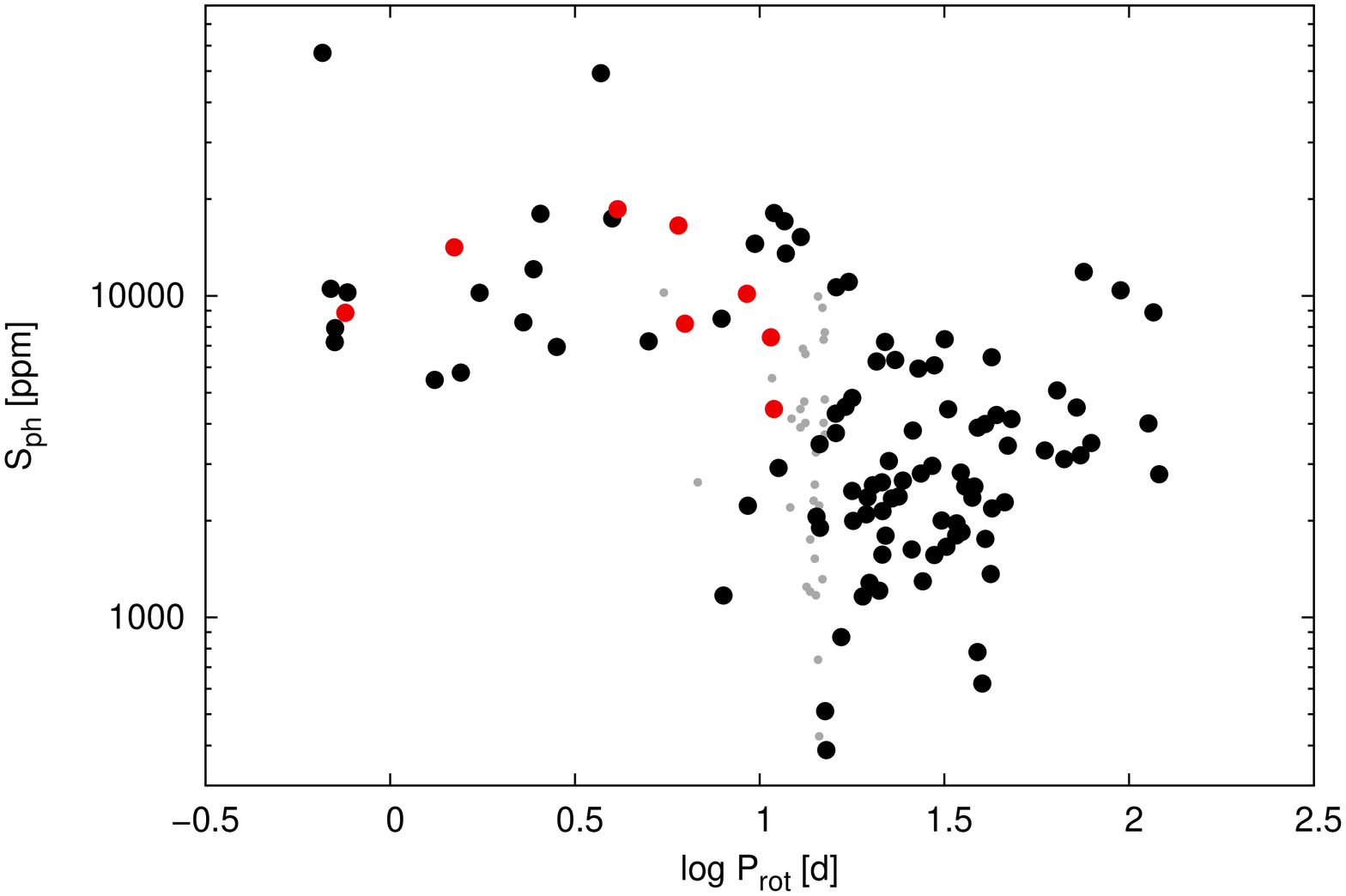} 
\includegraphics[width=6.5cm,angle=270]{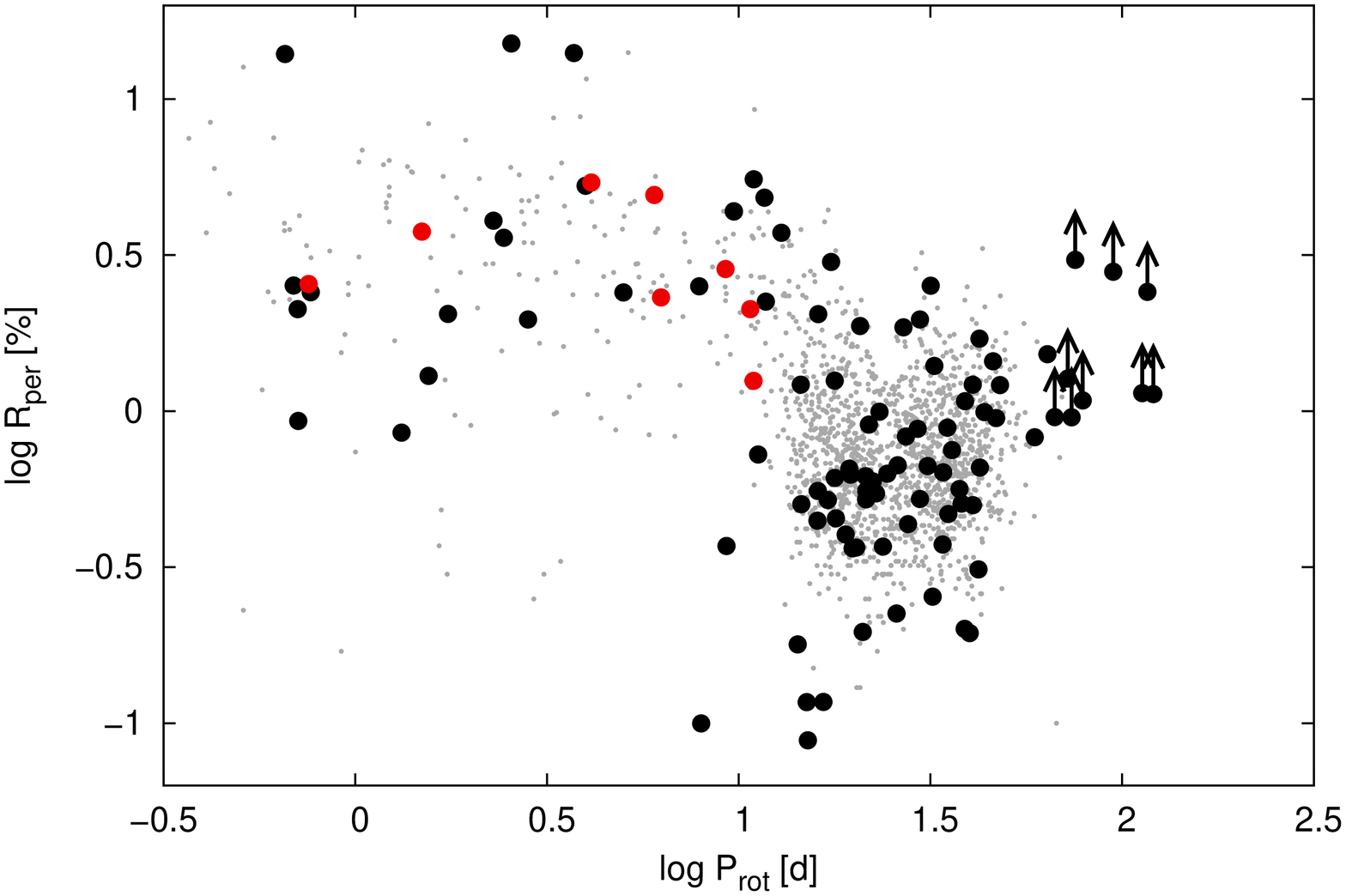} 
\caption{{Photometric activity diagnostics versus rotation period for M stars.  
Red dots -- M stars in our sample; 
black dots -- K2 M dwarf sample of \cite{2016MNRAS.tmp.1060S};
grey dots -- field M dwarfs
in the sample by \cite{2014JSWSC...4A..15M} in the top panel and
field M dwarfs from \cite{2013ApJ...775L..11M} in the bottom panel.
}}
\label{fig:sph_K1_K2} 
\end{center} 
\end{figure}
\indent We can examine the spectral type dependence of the photometric activity diagnostics, provided
that -- analogous to the detection of flares -- the brightness dependence of the 
sensitivity for measuring variations is considered.
In Fig.~\ref{fig:rper} we display $S_{\rm ph}$ and $R_{\rm per}$, 
versus $K_{\rm p}$ separately for each spectral class. 
The average values of $S_{\rm ph}$ and $R_{\rm per}$ 
for each spectral type are listed in Table~\ref{tab:mean_of_photact}. 
While the exclusively high values for the activity diagnostics in 
the M dwarfs may be due to their faintness, preventing the detection of small
variations, a curious distinction is seen for F stars. Albeit covering roughly the same
level of (quiescent) brightness in the {\em Kepler} band as the G dwarfs and some of the K dwarfs
and the same range of $P_{\rm rot}$ (see e.g. Fig.~\ref{fig:lx_prot}), 
the variability of the F stars is markedly smaller. This points at intrinsically weaker
variability in F stars. 
A similar effect is seen in Fig.4 of \cite{2014ApJS..211...24M}.
Our finding is aggravated by the fact that the F stars of our sample are
more active, i.e. have higher $S_{\rm ph}$ values, than a sample of $22$ F stars 
selected from the {\em Kepler} Asteroseismic Science Consortium (KASC) programme
studied by \cite{2014A&A...562A.124M}.}
\begin{figure} 
\begin{center}
\includegraphics[width=6.5cm,angle=270]{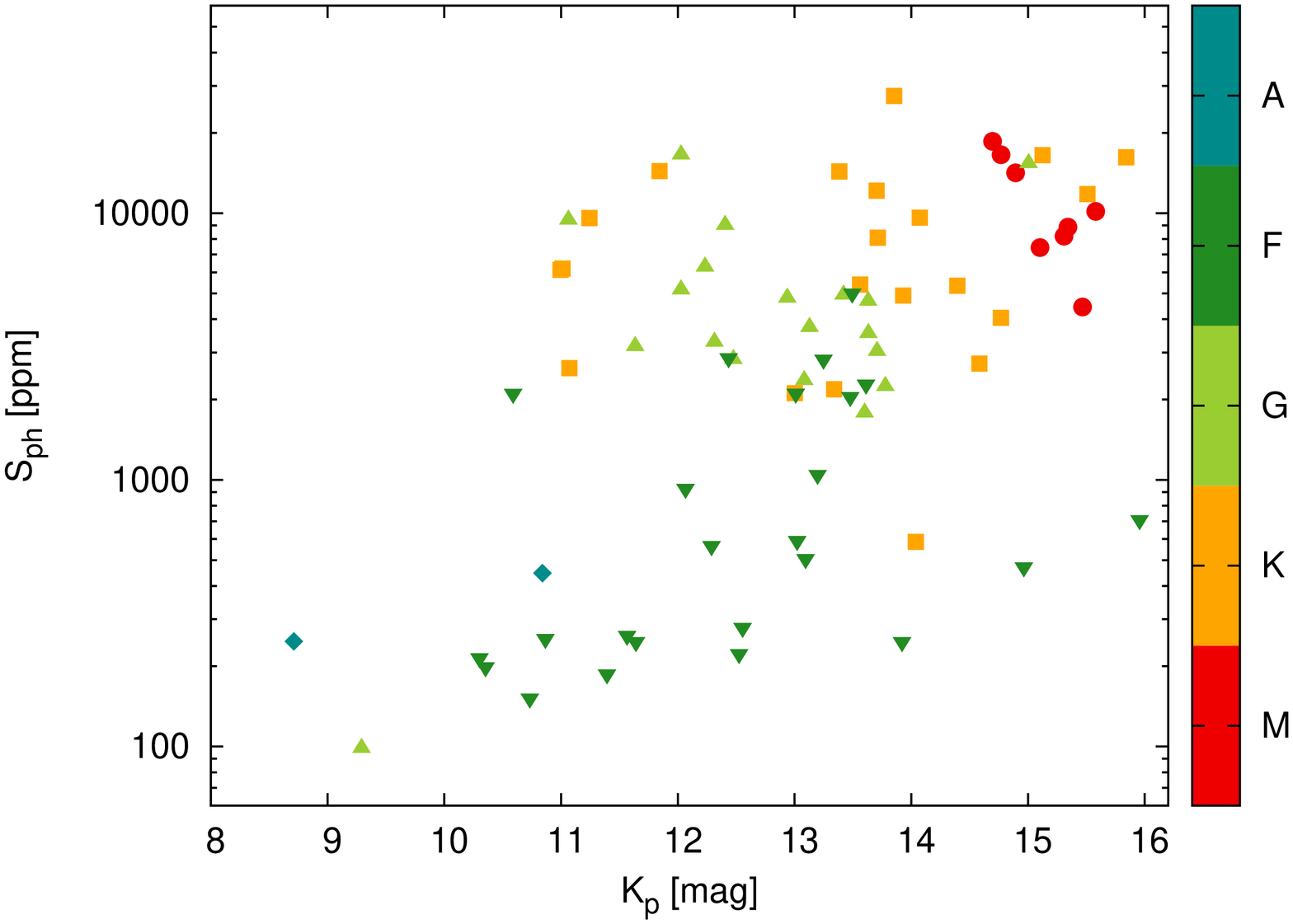} 
\includegraphics[width=6.5cm,angle=270]{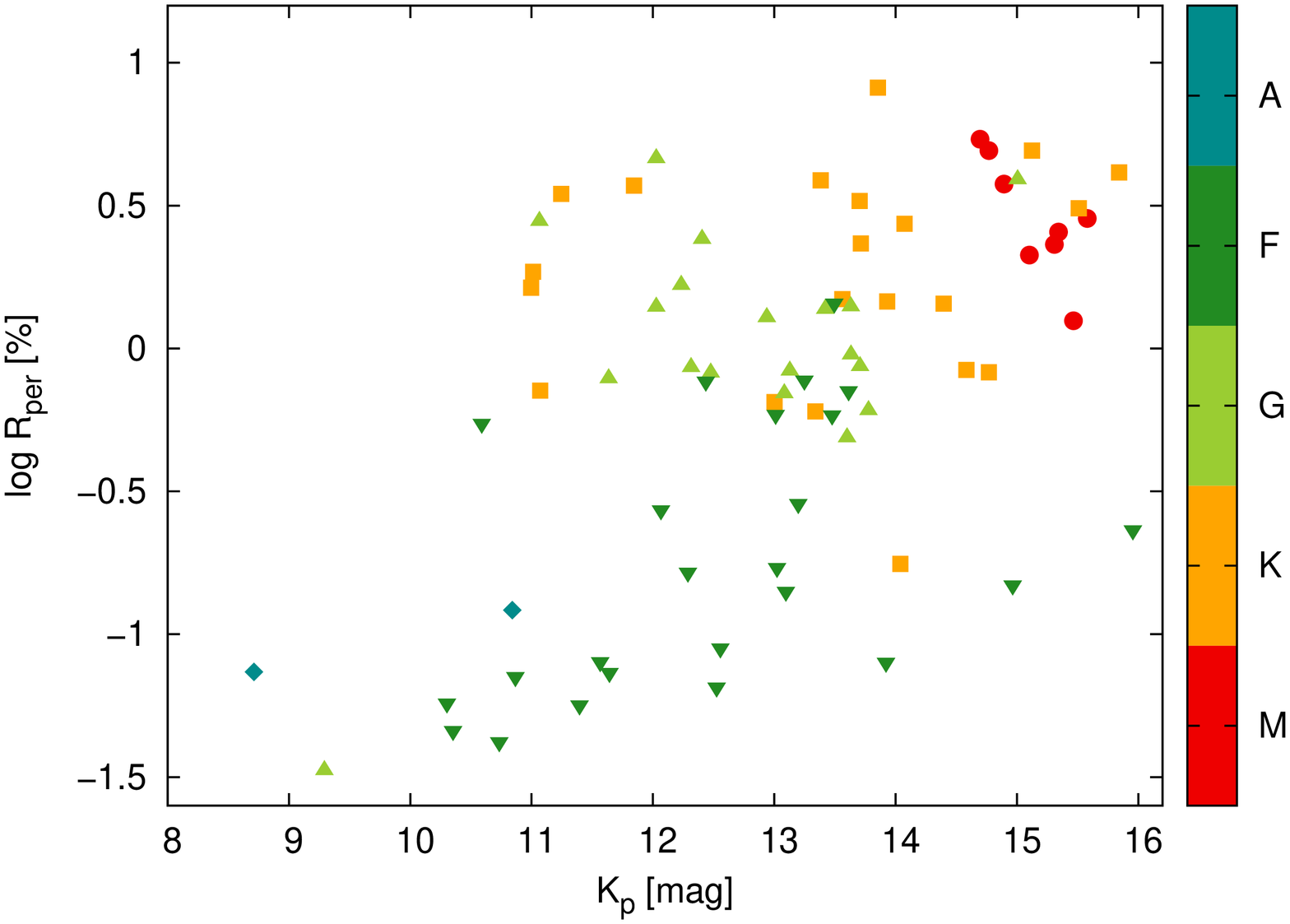} 
\caption{Photometric activity diagnostics versus {\em Kepler} magnitude
for the stars in our \kep / \xmm sample divided in spectral type bins. 
{\em Top}: $S_{ph}$ versus {\em Kepler} magnitude;  
{\em Bottom}: $\log{R_{\rm per}}$ versus {\em Kepler} magnitude.
} 
\label{fig:rper} 
\end{center} 
\end{figure}

\subsection{UV activity} \label{subsect:results_uv}


{As this sample is X-ray selected, all stars for which we detected an UV excess have
an X-ray detection. We show in Fig.~\ref{fig:lx_luv} the correlation between the 
UV excess luminosity (as calculated in Sect.~\ref{subsect:act_uv}) and the X-ray luminosity. 
With the caveat that the GALEX filters exclude 
some important chromospheric contributions, most notably the Mg\,{\sc ii} doublet, these relations
represent a comparison of the chromospheric and the coronal radiative energy output.
The sample with FUV excess is too small for any conclusion. We perform the Spearman's and Kendall's correlation 
tests for the NUV and X-ray luminosities of each SpT and find that only the K stars
show a significant positive correlation between the two luminosities (p-values $< 0.05$)
with rank correlation coefficients $\rho_{\rm s} = 0.85$ and $\rho_{\rm K} = 0.64$. 
The F and G stars show each a considerable spread in the distribution, 
similar to the behavior of the F stars in the X-ray / rotation 
relation of Fig.~\ref{fig:lx_prot}.}
\begin{figure*} 
\begin{center}
\parbox{16cm}{
\parbox{7.5cm}{
\includegraphics[width=6.5cm,angle=270]{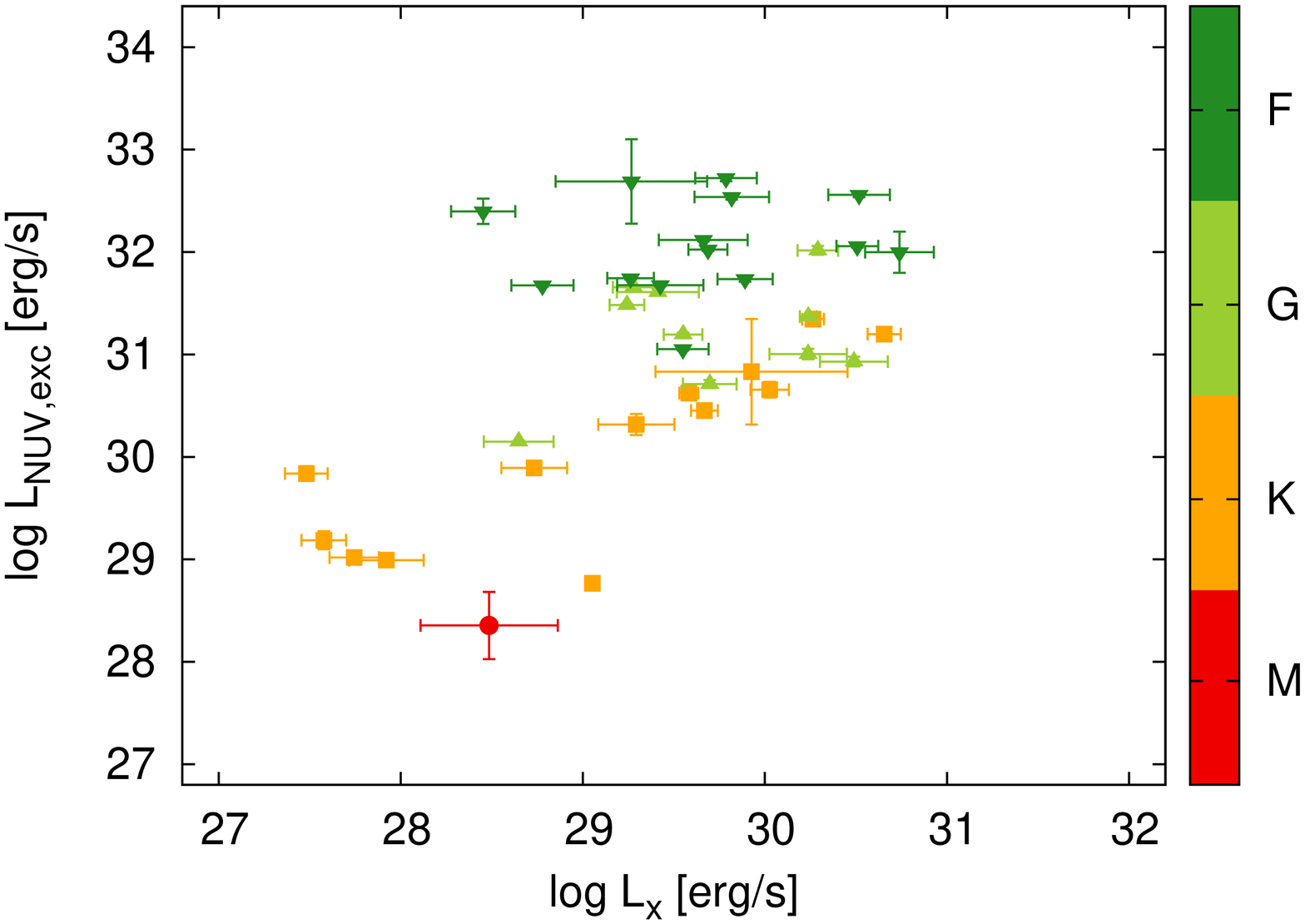}
}
\parbox{1cm}{
\hspace*{1cm}
}
\parbox{7.5cm}{
\includegraphics[width=6.5cm,angle=270]{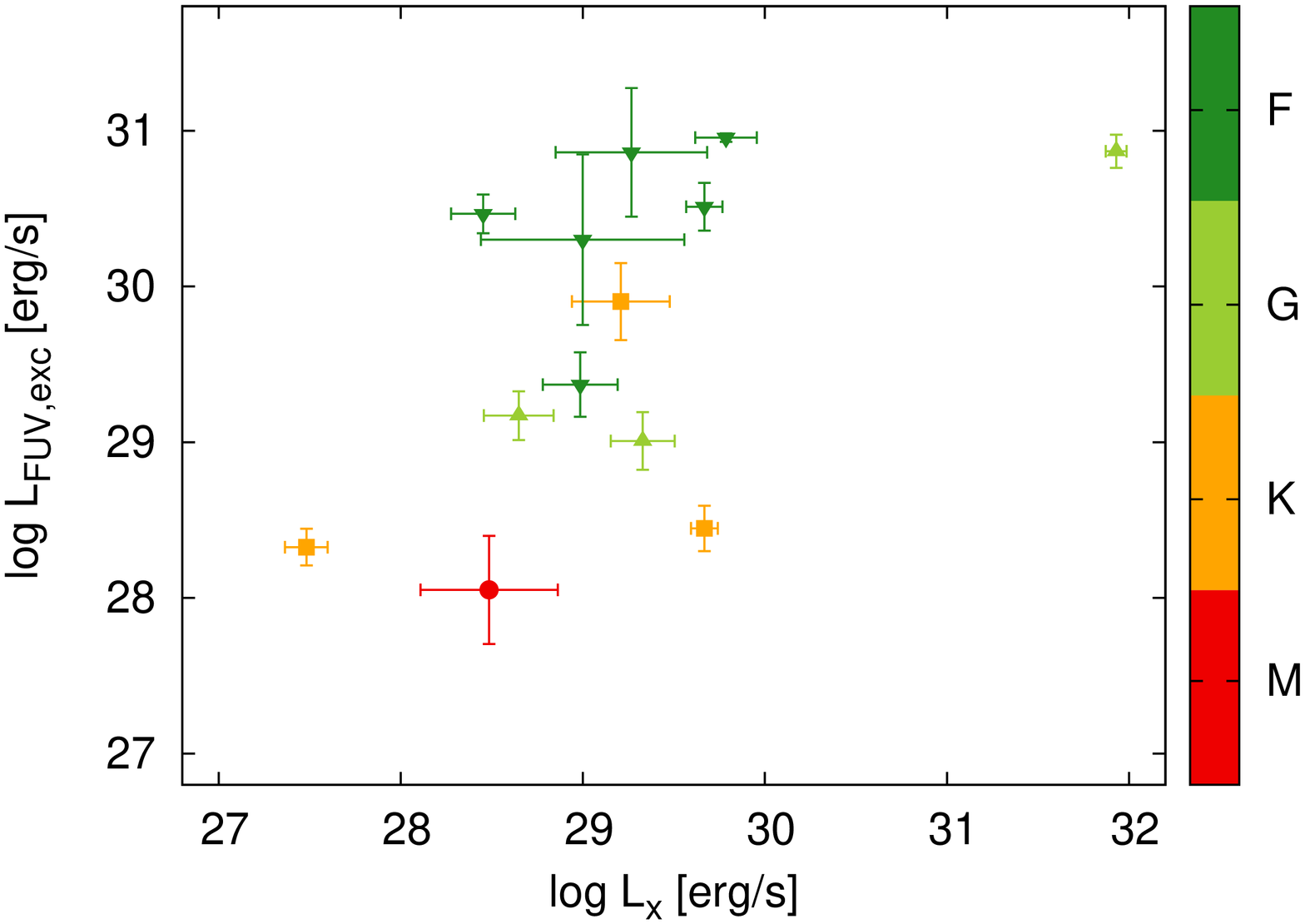}
}
}
\caption{{UV chromospheric excess luminosity versus X-ray luminosity for the full sample.}} 
\label{fig:lx_luv} 
\end{center} 
\end{figure*} 


{We also compare the UV emission levels of our sample to the UV
emission of the exoplanet hosts studied by \cite{2013ApJ...766....9S}. Deviating from the remainder
of this paper, which is focused on the high-energy emission due to stellar
activity, we consider here the total observed UV emission which includes photospheric plus 
chromospheric contribution. 
This enables a direct comparison to the work of \cite{2013ApJ...766....9S}. Secondly, 
for the irradiation of exoplanets the relevant parameter is the total UV output
of the star irrespective of its origin.}  
{Fig. \ref{fig:shkolnik} shows} the distribution of $L_{\rm NUV}/L_{\rm bol}$ versus 
$T_{\rm eff}$ for our stars and 
for the sample of exoplanet hosts analysed by \cite{2013ApJ...766....9S}. 
Our UV luminosity values are for given $T_{\rm eff}$ 
about one order of magnitude above the values of the stars with
exoplanets. This might be explained by an oppositely directed bias in both samples:
While our sample is biased towards active stars, {known} exoplanet hosts are typically
inactive as a result of the selection criteria for planet search programs. A second
reason that might influence the differences between the NUV emission of the two
samples is the fact that the stars in the sample of \cite{2013ApJ...766....9S} are
typically nearby (tens of parsec), such that many stars were saturated in 
the GALEX observations and they have been removed from the analysis, 
thus generating a bias towards small NUV luminosity. Based on these arguments,
there is good reason to believe that the two samples taken together represent the
full range of NUV luminosity for each spectral type. 

\begin{figure}
\begin{center}
\includegraphics[width=7.0cm,angle=270]{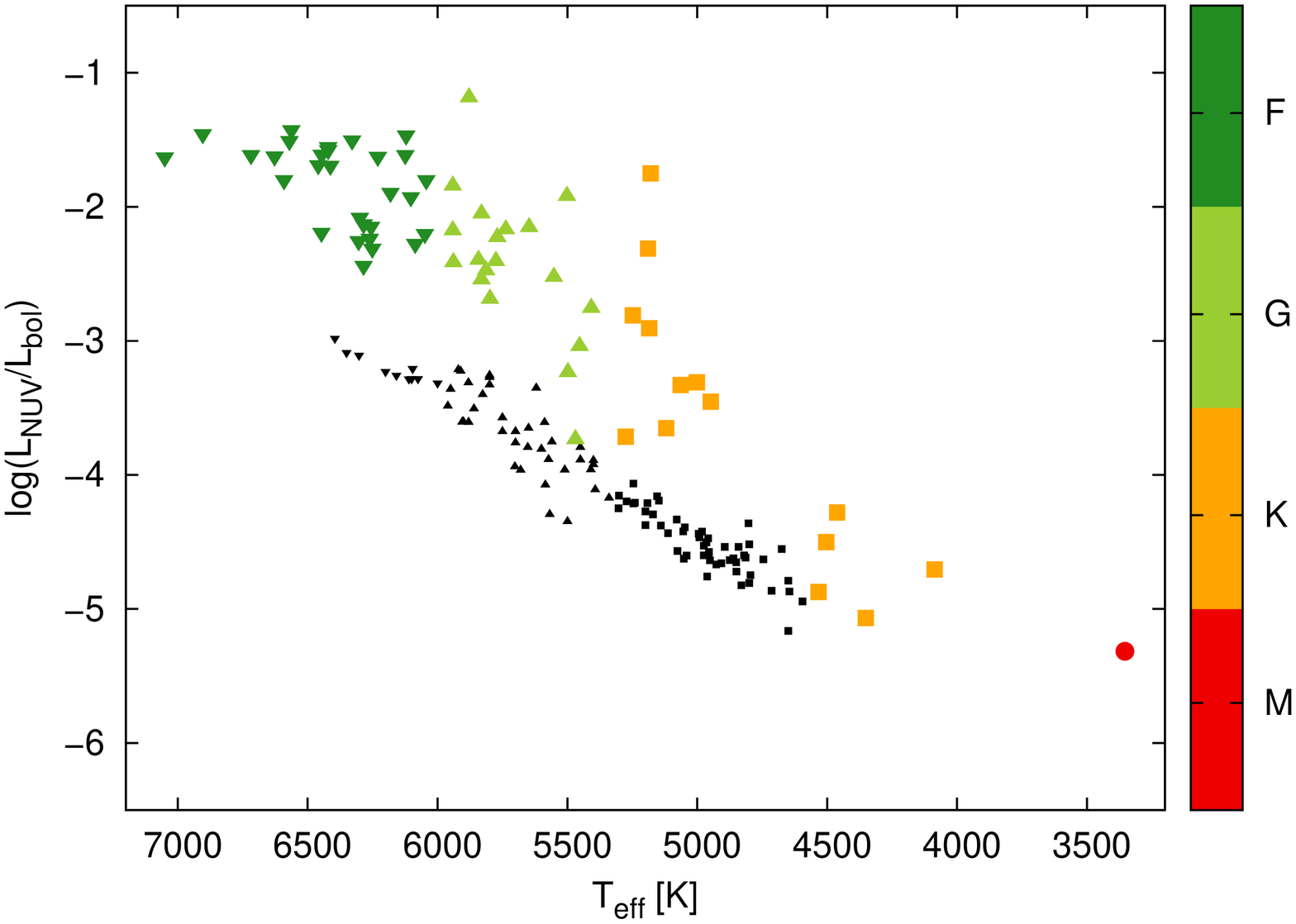}

\caption{{Observed fractional NUV luminosity} versus $T_{\rm eff}$ for the stars in our 
sample (with the usual colour code) and the ones from \cite[][black dots]{2013ApJ...766....9S}.	
}
\label{fig:shkolnik} \end{center} \end{figure}

\subsection{A-type stars} \label{subsect:results_A}

{\rm Three stars in our sample are classified as A-type} on the basis of their
effective temperature. The light curve of one of them (KIC\,5113797) shows a
variability pattern that suggests an association with the class of $\gamma\,$Dor
non-radial pulsators. The others (KIC\,8703413, KIC\,9048114) have a clear rotational
modulation pattern suggesting the presence of star spots. Only $3$ flares have been detected
by our algorithm in the \kep lightcurves of the A stars in our sample.  

Late A-type stars represent the high-mass borderline of the 
magnetically active stars. 
A change in the stellar structure occurs with respect to later-type stars: 
the outer convective layer disappears, leaving a fully radiative stellar interior. It
is generally believed that the standard stellar dynamo model can not be applied
to fully radiative stars, which lack the transition between convective and
radiative zones. Dynamos may operate in the small convective cores of hot stars, 
but a major problem is the emergence of the ensuing fields to the surface of the
stars where magnetic activity is observed. 
Considering these difficulties, the magnetic activity apparently
observed on A-type stars, and especially the X-ray emission, is often ascribed to
unresolved late-type companion stars in binary systems. 
In a systematic study of a large sample of A-type stars associated with a {\em ROSAT} 
X-ray source about $25$\,\% are bona-fide single stars \citep{2007A&A...475..677S}, and 
the question on whether A-type stars can maintain a corona has not been conclusively 
resolved. 

Similarly, the occurrence of photospheric magnetic activity in A-type stars is
disputed. 
\cite{2013MNRAS.431.2240B} and \cite{2015MNRAS.447.2714B} performed a search for
activity in the A stars observed by {\em Kepler}. Both rotational variability due to
starspots and flaring activity was detected for a significant fraction of stars.
{\rm Two A stars from our sample (KIC\,8703413 and KIC\,5113797) were classified as flaring stars
by \cite{2013MNRAS.431.2240B} and  \cite{2015MNRAS.447.2714B}.  
The authors considered the flares on {\em Kepler} A-type too energetic to be 
attributed to possible late-type companions.  
However, this result has been refuted by \cite{2016IAUS..320..150P}.} 

The $T_{\rm eff}$ of the three stars classified as A-type present in the \kep / \xmm sample is below the
limit set by \cite{2002ApJ...579..800S} for the fully radiative regime ($8250\,{K}$)
within uncertainties. KIC\,9048114 is also consistent, within uncertainties, with an
early-F type classification. The detection of activity from these stars {may thus not be}
inconsistent with the standard dynamo model.

\section{Conclusions} \label{sect:conclusions}

Using a complex sample selection approach involving multi-band photometric
characterisation and visual inspection of multi-band images, we identify 
$125$ stars 
with X-ray detection in the 3XMM-DR5 catalog that are observed by the {\em Kepler} mission.
The subsample of $102$ dwarf stars studied in this work  
comprises stars with spectral types from A to M. 
The distance of these stars range between $\sim \mathbf{40-7400}\,{\rm pc}$, with $90\,\%$ of the stars in the range $\sim\mathbf{40}-1500\,{\rm pc}$. 
By comparison to the volume-limited NEXXUS sample of nearby stars \citep{2004A&A...417..651S} 
we estimate that -- as a result of the X-ray selection and the large distances of the
{\em Kepler} stars -- we probe only the most active $10$\,\% of the stellar 
{main-sequence} population. 

A previous survey
of the X-ray (and UV) activity of \kep objects was performed by
\cite{2015AJ....150..126S}, in which $\sim 1/5$ of the \kep field of view was
surveyed with the {\em Swift} instruments XRT and UVOT (X-ray sensitivity in the
energy range $0.2-10\,{\rm keV}: 2\cdot10^{-14}\,{\rm erg\,cm^{-2}\,s^{-1}}$ in $10^4$\,s).
Ninety-three KIC objects were found {\rm to have} an X-ray counterpart, of which $33$
were recognised as stars. None of the $20$ X-ray emitting stars 
from our joint \kep / \xmm study that are in the {\em Swift} FoV was detected by
{\em Swift}. This is not surprising, as the {average} sensitivity limit of the {\em XMM-Newton}
observations in the {\em Kepler} field is {about a factor $3$} 
deeper than the {\em Swift} observations (see Table \ref{tab:obs} 
for the range of flux limits {for our data set}).  

In the \kep light curves of the XMM-DR5 selected sample, we found several
types of variability: rotational modulation due to spots, eclipsing binary systems,
multi-periodicity possibly associated with rotational modulation in binary systems,
contact binaries, variability due to stellar pulsations ($\delta\,$Sct and
$\gamma\,$Dor stars), uncorrelated and confuse variability patterns. A large fraction
of the stars 
present rotational modulation due to starspots ($74$,
$\sim73\%$ of the dwarf stars in the sample).
The number of stars with rotational modulation may represent a lower limit, since 
other stars may {have} a {weak} 
rotational modulation which cannot be ascertained with sufficient
confidence, or the rotational modulation is superimposed {on} other kinds of
variability, as {e.g. in the above-mentioned probable binary systems}.
The maximum period detectable is limited by the length of the \kep  
Quarters ($\sim 90\,{\rm d}$). We found periods in the
range $0.3-70\,{\rm d}$, with a distribution which does not present significant
differences from one spectral type to the other. 

We explored the relation of several photometric activity indicators and the coronal
X-ray emission with the rotation period. Our data confirm previously observed values
and ranges for the rotation-dependence of the star-spot brightness amplitude ($R_{\rm per}$), 
the overall variability (characterized by {the standard deviation of the lightcurve}  
$S_{\rm ph}$) and the residual noise after
subtraction of rotational modulation and flares ($S_{\rm flat}$). 
Due to the predominance of fast rotators in
our sample, the bimodality observed in previous works {on M dwarfs} 
\citep{2016MNRAS.tmp.1060S}, with a sharp transition between fast and slow
rotators, can not be studied at the same detail. We do find, however, 
{no evidence for high flare rates and amplitudes in the slow rotator regime above
$P_{\rm rot} \sim 10$\,d, in line with the expectation from these previous findings.} 

{The $R_{\rm per}$ and $S_{\rm ph}$ values of the F stars appear underluminous with
respect to later-type stars at the same intrinsic brightness, indicating that this result
is not a bias. The same effect was seen in larger and less biased samples, e.g. by
\cite{2014ApJS..211...24M}. Unresolved late-type companion stars are a possible explanation for the
low amplitude of the rotational modulation of the F stars. The rotational signal
might actually be from such an unknown late-type star, diluted by the bright F star
that dominates the unmodulated flux.
Alternatively, the low $R_{\rm per}$ and $S_{\rm ph}$ values of the F stars 
may mark decreasing spot filling factor towards the high-mass
limit of partially convective, dynamo-driving stars. 


{\rm To examine the relation between X-ray activity and rotation we {have composed}  
$L_{\rm X}$ vs $P_{\rm rot}$ and $L_{\rm X}/L_{\rm bol}$ vs Rossby number ($R_0$) {plot}. 
A kink in the rotation / X-ray relation is
observed, traditionally described as a transition between a saturated and a correlated
regime. 
The resulting distributions confirm earlier studies of this parameter space {using here} 
a much more homogeneous data set (X-ray data and rotation periods from a single 
instrument, each) and analysis (same flare and rotation search mechanism for all stars).
Similar to the most extensive sample studied previously in this respect, 
\cite{2011ApJ...743...48W}, the `saturated' and `correlated' regime 
are hard to define for subsamples divided by spectral
type. The reason is that in the representation involving Rossby number (where the
two regimes are usually more marked than when using $P_{\rm rot}$) individual spectral
types cover either only the saturated part (M stars) or only the correlated part
(FGK stars). 

We note that the F stars seem not to participate in the downward trend of the X-ray
emission in the $P_{\rm rot}$ (or $R_0$) range representing the `correlated' zone for the later
spectral types. {Also, contrary to the other spectral types, the F stars show no 
clear correlation between X-ray luminosity and UV excess luminosity. Within the group 
of F stars there is no correlation of X-ray or UV excess with $T_{\rm eff}$. It is  
therefore not obvious to ascribe this finding to the transition to fully radiative
interior with the ensuing break-down of the solar-type dynamo. One possibility for the
disordinated behavior of the F stars is that this group may comprise close binary stars with
a later type companion star confusing the activity pattern. In such unresolved binaries
the measured period might actually represent the rotation of the (unknown) later-type star
with implications on all relations involving $P_{\rm rot}$. In fact, we observe in 
Fig.~\ref{fig:lx_prot} that both the range of periods and the range of observed $L_{\rm x}$
are similar for M stars and F stars. This scenario would explain the absence of a
correlation between $P_{\rm rot}$ and $L_{\rm x}$ as the longest observed periods 
($\sim 10$\,d) are still within the range of the `canonical' saturated regime for M dwarfs.}  

A major asset of this study is the fact that $9$ of the
\xmm observations were carried out during the {\em Kepler} monitoring, providing strictly 
simultaneous optical and X-ray lightcurves. We discovered seven X-ray
flares. 
For all of them there is evidence of a white-light counterpart in the \kep lightcurves. 

\begin{acknowledgements}

This work is based on observations obtained with the {\it Kepler} mission and with 
{\em XMM-Newton}. Funding for the {\it Kepler} 
mission is provided by the NASA Science Mission directorate. {\em XMM-Newton} is an 
ESA science mission with 
instruments and contributions directly funded by ESA Member States and NASA. We have 
made use of data produced by the EXTraS project, funded by the European Union's 
Seventh Framework Programme, and we acknowledge its financial support under grant agreement 
no. 607452 (EXTraS project) and that of the Italian Space Agency (ASI) through the ASI-INAF 
agreement 2017-14-H.0. 
This work also includes data from the European Space Agency (ESA) mission
{\it Gaia} ({https://www.cosmos.esa.int/gaia}), processed by the {\it Gaia}
Data Processing and Analysis Consortium (DPAC,
{https://www.cosmos.esa.int/web/gaia/dpac/consortium}). Funding for the DPAC
has been provided by national institutions, in particular the institutions
participating in the {\it Gaia} Multilateral Agreement.
This research has made use of the SVO Filter Profile Service 
(http://svo2.cab.inta-csic.es/theory/fps/) supported from the Spanish MINECO through 
grant AyA2014-55216 and  of VOSA, developed under the Spanish Virtual Observatory 
project supported from the Spanish MICINN through grants AYA2008-02156 and AYA2011-24052. 

\end{acknowledgements}

\bibliography{keplerBS2.bib} \bibliographystyle{aa}

\onecolumn
\begin{landscape}
\renewcommand*{\arraystretch}{0.99}
\begin{longtable}{lccccccccccccr}
\caption{\label{tab:photometry}KIC \citep{2011AJ....142..112B} photometry {all} the 
$125$ stars in {the {\em Kepler}/ XMM-Newton} sample, with the corrections by {\cite{2012ApJS..199...30P}}
for the SLOAN filters and {updated magnitudes from the} UCAC-4 Catalogue \citep{2012yCat.1322....0Z}}\\
\hline 
KIC ID & $J$ & $H$ & $K_{\rm s}$ & $U$ & $B$ & $V$ & $K_{\rm p}$ & $g_{\rm SDSS}$ & $r_{\rm SDSS}$ & $i_{\rm SDSS}$ & 
$z_{\rm SDSS}$ & GALEX\_{\rm FUV}&GALEX\_{\rm NUV}\\
&{\rm [mag]}&{\rm [mag]}&{\rm [mag]}&{\rm [mag]}&{\rm [mag]}&{\rm [mag]}&{\rm [mag]}&{\rm [mag]}&
{\rm [mag]}&{\rm [mag]}&{\rm [mag]}&{\rm [mag]}&{\rm [mag]}\\ \hline
\endfirsthead
\caption{continued.}\\
\hline
KIC ID & $J$ & $H$ & $K_{\rm s}$ & $U$ & $B$ & $V$ & $K_{\rm p}$ & $g_{\rm SDSS}$ & $r_{\rm SDSS}$ & $i_{\rm SDSS}$ & 
$z_{\rm SDSS}$ & GALEX\_{\rm FUV}&GALEX\_{\rm NUV}\\
&{\rm [mag]}&{\rm [mag]}&{\rm [mag]}&{\rm [mag]}&{\rm [mag]}&{\rm [mag]}&{\rm [mag]}&{\rm [mag]}&
{\rm [mag]}&{\rm [mag]}&{\rm [mag]}&{\rm [mag]}&{\rm [mag]}\\ \hline
\endhead
\hline
\endfoot
4935950 & 9.257 & 8.679 & 8.513 & --- & --- & --- & 11.071 & 12.18 & 11.022 & 10.571 & 10.319 & --- & --- \\
4937350 & 13.109 & 12.756 & 12.698 & 15.291 & 15.222 & 14.587 & 14.273 & 14.664 & 14.197 & 14.034 & 13.946 & --- & --- \\
5023534 & 11.52 & 11.185 & 11.072 & --- & 13.758 & 13.061 & 13.083 & 13.688 & 13.013 & 12.767 & --- & --- & --- \\
5023985 & 12.412 & 12.125 & 12.082 & 14.489 & 14.346 & 13.715 & 13.495 & 13.858 & 13.42 & 13.267 & 13.224 & --- & --- \\
5024138 & 13.833 & 13.579 & 13.58 & 15.962 & 15.851 & 15.29 & 14.964 & 15.344 & 14.887 & 14.728 & 14.641 & --- & --- \\
5024146 & 13.573 & 13.24 & 13.162 & --- & --- & --- & 14.845 & 15.44 & 14.785 & 14.533 & 14.431 & --- & --- \\
5024450 & 13.942 & 13.628 & 13.56 & 15.976 & 15.856 & 15.284 & 15.063 & 15.476 & 14.981 & 14.816 & 14.719 & --- & --- \\
5025233 & 11.371 & 11.083 & 11.012 & 13.281 & 13.268 & 12.693 & 12.435 & 12.765 & 12.366 & 12.219 & 12.179 & --- & --- \\
\hline
\end{longtable}
\end{landscape}

\begin{landscape}
\renewcommand\tabcolsep{3pt}
\renewcommand*{\arraystretch}{0.99}
\begin{longtable}{lrrrrrcrrccl}
\caption{\label{table:params}Fundamental stellar parameters for all $125$ 
{ stars in the {\em Kepler} / {\em XMM-Newton} sample.}}\\ \hline
KIC ID & \multicolumn{1}{c}{$A_{\rm V}$} & \multicolumn{1}{c}{$\log{g_\odot}$} & \multicolumn{1}{c}{$[Fe/H]_\odot$} & \multicolumn{1}{c}{$T_{\rm eff}$} & \multicolumn{1}{c}{$d$} & Flag $d$ & \multicolumn{1}{c}{$M_{*}$} & \multicolumn{1}{c}{$\log{L_{\rm bol}}$} & SpT & Variability$^{(a)}$ & Flag \\
       & [mag]       &                 &                & \multicolumn{1}{c}{[K]}           & \multicolumn{1}{c}{[pc]} & & \multicolumn{1}{c}{[$M_\odot$]} &  \multicolumn{1}{c}{[$L_\odot$]}   &     &                     & \\  
\hline
\endfirsthead
\caption{continued.}\\
\hline
KIC ID & \multicolumn{1}{c}{$A_{\rm V}$} & \multicolumn{1}{c}{$\log{g_\odot}$} & \multicolumn{1}{c}{$[Fe/H]_\odot$} & \multicolumn{1}{c}{$T_{\rm eff}$} & \multicolumn{1}{c}{$d$} & Flag $d$ & \multicolumn{1}{c}{$M_{*}$} & \multicolumn{1}{c}{$\log{L_{\rm bol}}$} & SpT & Variability$^{(a)}$ & Flag \\
       & [mag]       &                 &                & \multicolumn{1}{c}{[K]}           & \multicolumn{1}{c}{[pc]} & & \multicolumn{1}{c}{[$M_\odot$]} &  \multicolumn{1}{c}{[$L_\odot$]}   &     &                     & \\  
\hline
\endhead
\hline
\endfoot
     4935950  & $0.04$ & $4.37$ & $0.040$ & $ 4381 \pm   124$ & $   64.7 \pm     6.1 $ &   Jmag  & $ 0.71 \pm  0.04$ & $-0.79 \pm  0.11 $ & K   & ROT    & \\
     4937350  & $0.44$ & $4.37$ & $-0.24$ & $ 6141 \pm   186$ & $ 1435.5 \pm    38.3 $ &   Gaia  & $ 1.13 \pm  0.13$ & $ 0.22 \pm  0.45 $ & F   & CONT?  & \\
     5023534  & $=0.38$ & $4.15$ & $-0.39$ & $ 5842 \pm   101$ & $  658.1 \pm    22.7 $ &   Jmag  & $ 0.93 \pm  0.00$ & $ 0.50 \pm  0.03$ & G   & ROT    & \\
     5023985  & $0.39$ & $4.27$ & $0.079$ & $ 6251 \pm   185$ & $  625.3 \pm     5.0 $ &   Gaia  & $ 1.18 \pm  0.22$ & $ 0.37 \pm  0.47 $ & F   & ROT    & \\
     5024138  & $0.54$ & $4.48$ & $-1.90$ & $ 6156 \pm   204$ & $  774.4 \pm    66.4 $ &   Jmag  & $ 0.75 \pm  0.03$ & $-0.21 \pm  0.11 $ & F   & ROT    & \\
     5024146  & $0.57$ & $4.24$ & $-1.53$ & $ 6121 \pm   214$ & $  772.6 \pm    98.8 $ &   Jmag  & $ 0.74 \pm  0.03$ & $-0.12 \pm  0.14 $ & F   & ECL    & \\
     5024450  & $0.54$ & $4.42$ & $-0.06$ & $ 6199 \pm   216$ & $ 2551.9 \pm   145.3 $ &   Gaia  & $ 1.24 \pm  0.15$ & $ 0.36 \pm  0.29 $ & F   & ECL    & \\
     5025233  & $0.25$ & $4.30$ & $-0.40$ & $ 6103 \pm   168$ & $  401.7 \pm   183.5 $ &   Jmag  & $ 0.89 \pm  0.13$ & $ 0.18 \pm  0.46 $ & F   & ROT    & \\
\end{longtable}
\end{landscape}

\begin{landscape}
\begin{longtable}{rrrrrrrrrrr}
\caption{\label{tab:rot}Rotation and flaring parameters derived from the \kep light curves for all $74$ rotationally-variable main-sequence stars in { the {\em Kepler} / {\em XMM-Newton}} sample} \\ 
\hline
KIC ID & $P_{\rm rot}$ & $R_{\rm 0}$ & $\log{R_{\rm per}}$ & $S_{\rm ph}$ & $S_{\rm ph,5}$ & $S_{\rm flat} $ & $<\log{A_{\rm peak}}>$ & $\log{A_{\rm peak, min}}$ & $\log{A_{\rm peak, max}}$ & Flare rate \\ 
       & [d]           &             & $[\%]$        & [ppm]        & [ppm]          &[ppm]           &                    &                            &                            & $[{\rm d^{-1}}]$ \\
  & (1) & (2) & (3) & (4) & (5) & (6) & (7) & (8) & (9) & (10)  \\ \hline
\endfirsthead
\caption{continued.} \\
\hline
KIC ID & $P_{\rm rot}$ & $R_{\rm 0}$ & $\log{R_{\rm per}}$ & $S_{\rm ph}$ & $S_{\rm ph,5}$ & $S_{\rm flat} $ & $<\log{A_{\rm peak}}>$ & $\log{A_{\rm peak, min}}$ & $\log{A_{\rm peak, max}}$ & Flare rate \\ 
       & [d]           &             & $[\%]$        & [ppm]        & [ppm]          &[ppm]           &                    &                            &                            & $[{\rm d^{-1}}]$ \\
  & (1) & (2) & (3) & (4) & (5) & (6) & (7) & (8) & (9) & (10)  \\ \hline
\endhead
\hline
\endfoot
     4935950 & $ 12.66$ & $  0.38$ & $  -0.15$ & $      2621$ & $      2467$ & $    41.9$ & $ -3.28$ & $ -3.56$ & $ -2.34$ & $  0.012 $ \\
     5023534 & $ 17.76$ & $  1.52$ & $  -0.16$ & $      2361$ & $      2393$ & $   109.4$ & $ -2.94$ & $ -3.17$ & $ -2.48$ & $  0.015 $ \\
     5023985 & $  6.58$ & $  1.40$ & $   0.16$ & $      5004$ & $      4911$ & $   142.0$ & $ -2.82$ & $ -2.83$ & $ -2.46$ & $  0.003 $ \\
     5024138 & $  7.52$ & $  1.20$ & $  -0.83$ & $       471$ & $       467$ & $   326.4$ & $         -$ & $         -$ & $         -$ & $         - $ \\
     5025233 & $  4.04$ & $  0.56$ & $  -0.11$ & $      2866$ & $      2799$ & $    93.4$ & $ -2.83$ & $ -3.33$ & $ -2.01$ & $  0.060 $ \\
     5111932 & $  1.45$ & $  1.56$ & $  -0.78$ & $       565$ & $       542$ & $    79.0$ & $ -3.07$ & $ -3.34$ & $ -2.79$ & $  0.015 $ \\
     5112483 & $ 63.51$ & $  1.64$ & $   0.54$ & $      9582$ & $         0$ & $    55.2$ & $ -3.25$ & $ -3.58$ & $ -2.93$ & $  0.006 $ \\
     5112508 & $  0.76$ & $  0.02$ & $   0.41$ & $      8858$ & $      8309$ & $  1021.9$ & $ -1.49$ & $ -2.17$ & $ -1.09$ & $  0.098 $ \\
\end{longtable}
\end{landscape}

\renewcommand*{\arraystretch}{0.99}
\begin{longtable}{rccccc}
\caption{\label{tab:xray} X-ray luminosity (in the $0.2-2$\,keV energy band), observed NUV and FUV luminosity, and NUV and FUV chromospheric excess luminosity.}\\ \hline
 KIC\_ID & $L_{\rm X}$ & $L_{\rm NUV}$ & $L_{\rm FUV}$ & $L_{\rm NUV,exc}$ & $L_{\rm FUV,exc}$ \\
         & [$10^{29}$\,erg/s] & [$10^{29}$\,erg/s] & [$10^{29}$\,erg/s] & [$10^{29}$\,erg/s] & [$10^{29}$\,erg/s] \\ \hline
\endfirsthead
\caption{continued.}\\ \hline
 KIC\_ID & $L_{\rm X}$ & $L_{\rm NUV}$ & $L_{\rm FUV}$ & $L_{\rm NUV,exc}$ & $L_{\rm FUV,exc}$ \\
         & [$10^{29}$\,erg/s] & [$10^{29}$\,erg/s] & [$10^{29}$\,erg/s] & [$10^{29}$\,erg/s] & [$10^{29}$\,erg/s] \\ \hline
\endhead
\hline
\endfoot
     4935950 & $      0.04  \pm       0.02$ & $         -                $ & $           -                 $ & $         -$ & $           - $ \\
     4937350 & $     21.40  \pm       9.08$ & $         -                $ & $           -                 $ & $         -$ & $           - $ \\
     5023534 & $      2.58  \pm       1.34$ & $     896.0   \pm      63.3$ & $           -                 $ & $     406.0$ & $         - $ \\
     5023985 & $      1.96  \pm       0.99$ & $     490.0   \pm      13.4$ & $           -                 $ & $         -$ & $         - $ \\
     5024138 & $      2.52  \pm       1.29$ & $         -                $ & $           -                 $ & $         -$ & $           - $ \\
     5024146 & $      4.51  \pm       1.91$ & $     175.0   \pm      47.6$ & $           -                 $ & $         -$ & $         - $ \\
     5024450 & $     40.05  \pm      15.43$ & $         -                $ & $           -                 $ & $         -$ & $           - $ \\
     5025233 & $      4.07  \pm       3.91$ & $     674.0   \pm     615.0$ & $           -                 $ & $         -$ & $         - $ \\
\hline
\end{longtable}

\end{document}